\documentclass{aa}
\usepackage[varg]{txfonts}
\usepackage{natbib}
\usepackage{graphicx}
\usepackage{color}
\usepackage{blindtext, subfig}
\usepackage{dblfloatfix} 
\usepackage{tikz}
\usepackage{longtable}
\usepackage{soul,xcolor} 

\graphicspath{{./}{./Figures/}}

\newcommand{\etal}{ et al. }
\newcommand{\be}{\begin{equation}}
\newcommand{\ee}{\end{equation}}
\newcommand{\beq}{\begin{eqnarray}}
\newcommand{\eeq}{\end{eqnarray}}

\newcommand{\todash}{\,--\,}

\usetikzlibrary{arrows,positioning} 
\tikzset{
    >=stealth',
    punkt/.style={
           rectangle,
           rounded corners,
           draw=black, very thick,
           text width=6.5em,
           minimum height=2em,
           text centered},
    pil/.style={
           ->,
           thick,
           shorten <=2pt,
           shorten >=2pt,}
}

\begin{document}

\setstcolor{red}   

\title{Properties of solar energetic particle events inferred from their associated radio emission}

\author{A. Kouloumvakos\inst{1} \and A. Nindos\inst{1} \and  E. Valtonen\inst{2} \and C.E. Alissandrakis\inst{1} \and O. Malandraki\inst{3} \and \\ P. Tsitsipis\inst{4} \and  A. Kontogeorgos\inst{4} \and X. Moussas\inst{5} \and A. Hillaris\inst{5}  } 

\offprints{A. Kouloumvakos, \email{akouloumvak@phys.uoa.gr}}

\institute{Section of Astrogeophysics,  Department of Physics, University of Ioannina, Greece.
\and
Department of Physics and Astronomy, University of Turku, 20014 Finland.
\and
IAASARS, National Observatory of Athens, GR-15236, Penteli, Greece
\and
Department of Electronics, Technological Educational Institute of Lamia, Greece.
\and
Section of Astrophysics, Astronomy and Mechanics, Department of Physics,
National and Capodistrian University of Athens, Greece.} 

\abstract
{}{We study selected properties of solar energetic particle (SEP) events as inferred from their associated radio emissions.}
{We used a catalogue of 115 SEP events, which consists of entries of proton intensity enhancements at one AU, with complete coverage over solar cycle 23 based on high-energy ($\sim$\,68~MeV) protons from SOHO/ERNE. We also calculated the proton release time at the Sun using velocity dispersion analysis (VDA). After an initial rejection of cases with unrealistic VDA path lengths, we assembled composite radio spectra for the remaining events using data from ground-based and space-borne radio spectrographs.  We registered the associated radio emissions for every event, and we divided the events in groups according to their associated radio emissions. In cases of type III-associated events, we extended our study to the timings between the type III radio emission, the proton release, and the electron release as inferred from VDA based on Wind/3DP 20--646 keV data.} 
{The proton release was found to be most often accompanied by both type III and II radio bursts, but a good association percentage was also registered in cases accompanied by type IIIs only. The worst association was found for the cases only associated with type II. In the type III-associated cases, we usually found systematic delays of both the proton and electron release times as inferred by the particles' VDAs, with respect to the start of the associated type III burst. The comparison of the proton and electron release times revealed that, in more than half of the cases, the protons and electrons were simultaneously released within the statistical uncertainty of our analysis. For the cases with type II radio association, we found that the distribution of the proton release heights had a maximum at $\sim$2.5\,$\mathrm{R_\odot}$. Most (69$\%$) of the flares associated with our SEP events were located in the western hemisphere, with a peak within the well-connected region of $50^\circ$\,--\,$60^\circ$ western longitude.}
{Both flare- and shock-related particle release processes are observed in major proton events at $>$50 MeV. Typically, the protons are released after the start of the associated type III bursts and simultaneously or before the release of energetic electrons. Our study indicates that a clear-cut distinction between flare-related and CME-related SEP events is difficult to establish. } 

\date{Received ........ / Accepted .......}

\keywords{Sun: particle emission - Sun: radio radiation}
\authorrunning{Kouloumvakos \etal}
\titlerunning{Radio Emission of SEP Events}
\maketitle

\section{Introduction}\label{sec:intro}

Solar energetic particle (SEP) events, are among the most important elements of space weather \citep[\textit{e.g.}][]{Reames1999,Vainio2009,Gopalswamy2008,Valtonen2011}. The energy of SEPs usually range from $\sim$10 keV\,/\,nucleon up to some GeV, the events last from several hours to a few days, and their fluxes can rise above the background by many orders of magnitude. They accompany solar flares and coronal mass ejections (CMEs) \citep[\textit{cf.} reviews by][]{Reames1999,Cliver2000}, and both types of activity provide possibilities for SEP-related particle acceleration through magnetic reconnection and large-scale coronal disturbances such as shock waves. An important issue is whether both magnetic reconnection and shock wave acceleration can simultaneously accelerate particles or whether one of them dominates the particles energization process. The scenario of diffusive shock acceleration seems to be a plausible source of low-energy protons up to some MeV; however, it is unlikely that the shock mechanism alone applies to very high-energy particles from the Sun, in the relativistic range.

Work in the 1990s indicated that the SEP events could be separated into two distinct classes based on the properties of the associated SXR flare, correlations with radio bursts, the presence or absence of a CME, the abundances, and charge states. The SEP events were classified as either impulsive or gradual \citep[\textit{cf.}][]{Reames1999,Kallenrode2003,Reames2013}. \textit{\textup{Impulsive}} SEP events may originate from solar flares, usually small flares \citep[C-M-class; see][]{Nitta2006}, last less than one day, have relatively small peak fluxes, and have been associated with type III radio bursts \citep{Reames1986}. \textit{\textup{Gradual}} SEP events are thought to be accelerated in CME-driven coronal and/or interplanetary shocks. Compared to impulsive SEP events, gradual SEPs have long durations (\textit{i.e.}\,days), relatively high peak fluxes, and have long been associated with type II radio bursts \citep{Lin1970,Svestka1974,Cliver2004}. Although this classification has been useful, we point out that both flares and CMEs are associated with almost all SEP events, so it is often difficult to distinguish the particle accelerator unambiguously.

Radio emission in the solar corona and interplanetary (IP) medium provides important information about the acceleration and propagation of solar energetic particles as well as shocks.  The different types of radio bursts in the dynamic radio spectrum, are usually interpreted as signatures of electron beams \citep[type IIIs:][]{Lin1981,Lin1986}, electrons accelerated in shock waves \citep[type IIs,][]{Nelson1985,Cairns2003,Ganse2012}, or as electrons confined (type IVs) in closed loop structures \citep[see \textit{e.g.}][]{Nindos2008}. 

So far various studies have investigated the association between the type III, II, and IV radio bursts and SEP events in general. \citet{Lin1970} found that SEP events from 1964 to 1967 were mostly accompanied (70$\%$) by type III and II radio bursts; subsequently \citet{Svestka1974}, using type II burst records and SEPs from 1966 to 1968, concluded that almost all type II radio bursts were associated with solar proton events. \citet{Kahler1982} suggested that only about half of metric type II bursts with sources in the western solar hemisphere were linked to SEPs. Even though coronal and IP shock waves are well established to be the principal accelerators for the largest observed SEP events, not all coronal shocks are "SEP effective" \citep[see][]{Gopalswamy2008}. \citet{Cane2002}  concluded that the proton events detected above $\sim$20 MeV (Jan. 1997\todash{}May 2001) were always accompanied by type III bursts, by CMEs and, for events originating not too far behind the limb, by flare signatures in $\mathrm{H\alpha}$ and soft X-rays. Even though \citet{Cane2002} reported that all the SEP events are accompanied by CMEs, their study focussed only on type III radio association and a possible type II occurrence was not presented.

 The statistics presented in \citet{Cliver2004}, from 1996 July through 2001 June, supported the widely held view that the SEPs ($\sim$20 MeV) observed near Earth were associated with coronal shock waves and especially with metric type II bursts that have decametric-hectometric (DH) counterparts. \citet{Cliver2009} separated their SEP sample into gradual and impulsive events and found that gradual events are always accompanied by DH-type II radio bursts, impulsive events had no association with DH-type II (except for one case), and in all cases an associated type III burst was observed. Also type III bursts with associated DH-type II bursts were ten times more likely to have large associated SEP events above 30 MeV \citep[also see][]{Gopalswamy2010}. In the most recent study presented by \citet{Miteva2013}, SEP events, either gradual or impulsive, were found to have the highest association rate with type III radio bursts and a lower association with type II bursts.
 
The comparison of the release times, between ions, electrons, and electromagnetic emission can, in principle, be used to identify the most relevant acceleration processes. Radio waves originate from energetic electrons, thus any comparison with ions in space is indirect. So far, energetic electrons have been studied with respect to their associated radio emissions and especially with type III radio bursts \citep[\textit{e.g.}][]{Haggerty2002,Cane2003}. \citet{Krucker1999} and \citet{Haggerty2002} concluded that the release of electrons into interplanetary space often starts several minutes after the first radiative signatures of electrons in the corona, assuming negligible interplanetary propagation effects. The late electron release can be ascribed to: (1) the action of a different acceleration process for the escaping electrons, \textit{i.e.} the delay is due to electron acceleration at a coronal shock \citep{Simnett2002}; (2)  transport effects during interplanetary propagation \citep{Cane2003}; (3) delayed electrons  originating at a reconnection site behind the CME shock front at regions of magnetic restructuring linked to the CME development \citep[see][]{Maia2007}. 

Compared with proton release time, relativistic electrons were found to be released with a significant delay in many cases \citep{Cliver1982}. \citet{Cliver1982} concluded that 100 keV electrons were released first, followed by the 2 GeV protons within 5 minutes, and 1 MeV electrons following at least 5 minutes later. This apparent delay in proton-electron acceleration may be due to several causes, such as (1) selective acceleration, (2) cross-field transport or perpendicular particle diffusion, and (3) particle trapping.

In this article we study selected properties of high-energy SEP events, especially focussing on the association of the SEP release time, as inferred by the velocity dispersion analysis (VDA), with transient solar radio emissions recorded by space and ground-based radio spectrographs. We have also thoroughly examined the time difference of proton release with respect to the escape of keV electrons into space, derived from both the electron release as traced by DH type III bursts and the electron release as derived from VDA. For the cases with type II radio association, we estimated the proton release heights from the height-time profile of the CME leading edge.


\section{Data selection and analysis}\label{sec:Data}
\subsection{Proton event selection}
\begin{figure}[!t]
\begin{center}
\centerline{\includegraphics[width=0.49\textwidth]{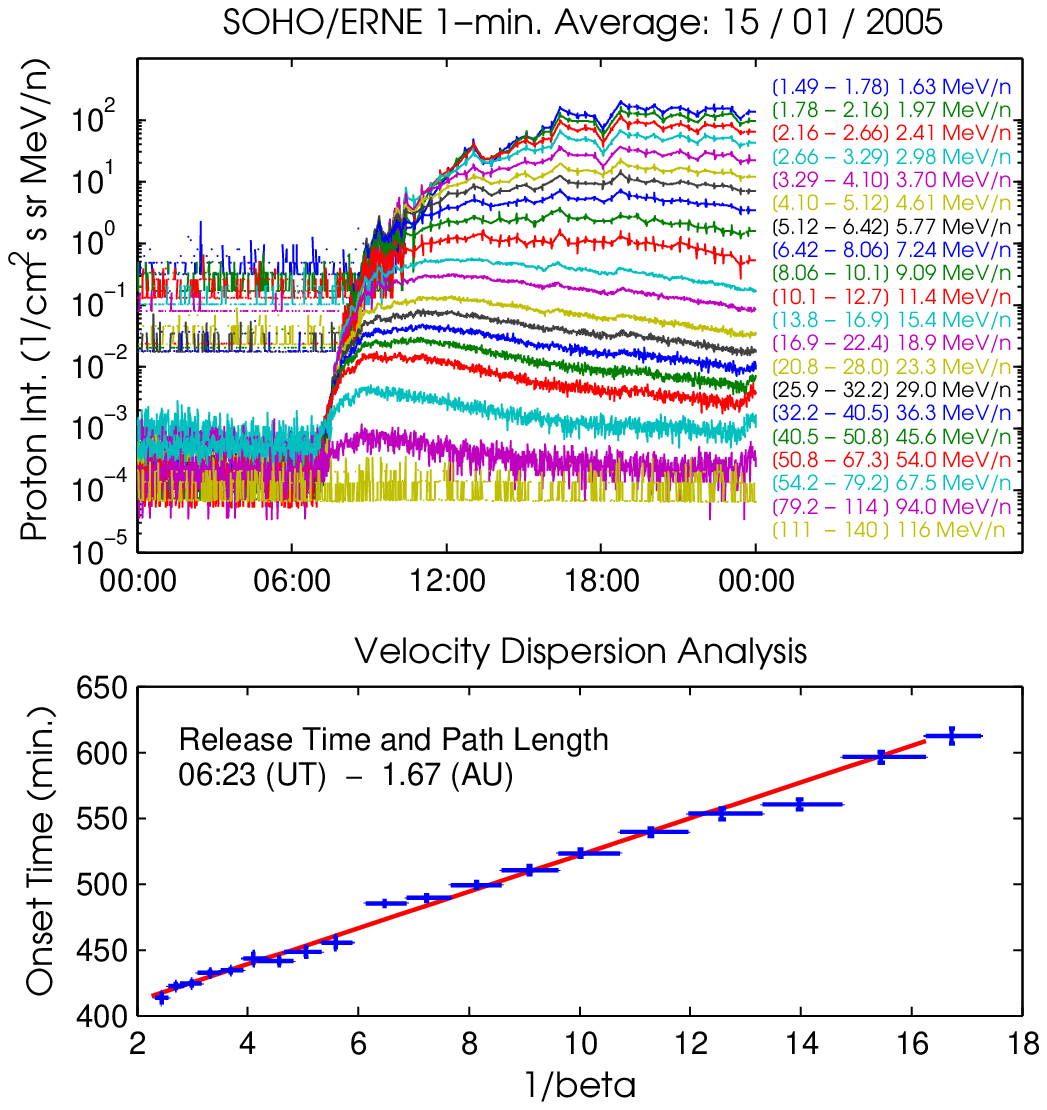}} 
\centerline{\includegraphics[width=0.5\textwidth]{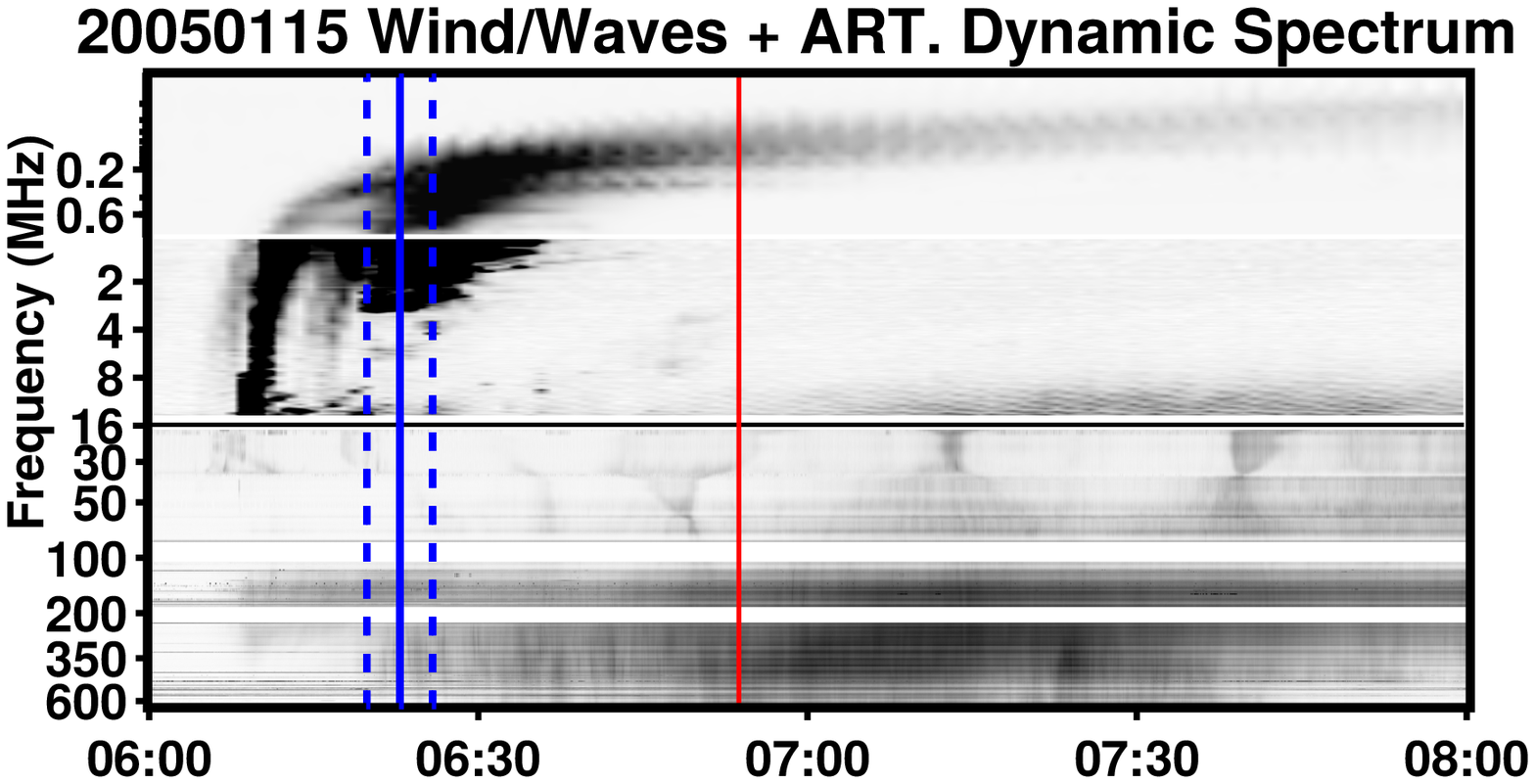}}
\vspace{-0.028\textwidth}
\end{center}
   \caption{Proton Event of 15 January 2005: 
   Top: proton intensity as recorded from SOHO/ERNE for 20 proton energy channels between 1.58 MeV and 131 MeV. 
   Middle: proton velocity dispersion analysis from the SOHO/ERNE data. We mark the proton onset times in the individual energy channels with blue crosses, the horizontal error bar represents the uncertainty of the particles' inverse velocity, which is determined from the width of every energy channel and the vertical error bar represents the uncertainty of the onset time determination derived from the Monte Carlo analysis. With the straight red line we present the linear fit to the onset times using the maximum likelihood estimation approach.
   Bottom: composite radio spectra of Wind/Waves and ARTEMIS-IV. The vertical blue solid line and dashed lines denote the proton release time derived from the VDA and its uncertainty, respectively;  the red line denotes the onset time of 54.8--80.3 MeV protons.} 
         \label{Fig:VDAprev}
   \end{figure}

For the purpose of our analysis, we used a list of 115 high-energy solar proton events as recorded by the \textit{Energetic and Relativistic Nuclei and Electron} \citep[ERNE:\,][]{Torsti1995} instrument of the University of Turku, on the \textit{Solar and Heliospheric Observatory} (SOHO) from May 1996 until the end of 2010 extending over a full solar cycle. The catalogue of these SEP events was presented in \citet{Vainio2013} and consists of entries of proton intensity enhancements based on the systematic scan of proton intensities in the energy range of 54.8\,--\,80.3 MeV. To register a proton intensity enhancement as a SEP event, we used the criterion that the 1 minute average intensity in the 54.8\,--\,80.3 MeV ERNE proton channel should be enhanced by a factor of $\sim$\,3 above the quiet time background. In Fig.~\ref{Fig:VDAprev} we present a representative case, the 15 January 2005 SEP event.

The energy range of 54.8\,--\,80.3 MeV in the SEP selection criterion was purposely chosen well above the typically used >10 MeV proton channel \citep[\textit{e.g.}][]{Laurenza2009}, available from GOES satellites, to identify individual events in cases in which large SEP events mask smaller events at low energies when they follow each other in quick succession. At high energies these cases can be better distinguished because of a more rapid fall of the proton intensities. However, the ERNE High Energy Detector has a very large geometric factor \citep[$\sim$\,$20$\,--\,$40~\mathrm{cm^2\,sr}$;][]{Valtonen1997,Torsti1995} often causing saturation during very large SEP events. Therefore, events following larger events before the recovery of the instrument may be missed. These cases, among others, are the events of 17 and 20 January 2005 following the event of 15 January 2005  \citep[see Table 1 of][list of SEP events observed from 1997 to 2006, at energies above 25 MeV]{Cane2010}, which were not included in our catalogue. 

\subsection{Estimate of proton release time} \label{Subsect:ProtRelTime}

 We determined the proton onset time with the so-called Poisson-CUSUM method in 20 proton energy channels of SOHO/ERNE between 1.58\,MeV and 131\,MeV for each SEP event. Cumulative sum (CUSUM) quality control schemes were proposed by \citet{Page1954} and are widely used in other fields, especially in medical research. A CUSUM scheme cumulates the difference between an observed and a reference value. If this cumulation equals or exceeds a threshold, then an out-of-control signal is given at the exact moment when the process transition to the observed value has occurred. 
 
 Based on the definition of the CUSUM schemes, the onset time of an event, at a given energy channel, is the instant when the signal has a systematic change of its mean value compared to the pre-event background mean. Usually, to consider that a transition in the signal mean value has occurred we use the two-sigma-shift criterion \citep{Huttunen2005,Vainio2013}. Previous studies have used other methods for the determination of the particles' onset times: for example \cite{Krucker1999} used a Shewhart control-chart scheme with an upper control limit at $4\,\sigma$ and \cite{Masson2012} presented a method in which the onset times were determined by reference times in the background-normalized rise time profiles at different energies, which works similar to a cross-correlation analysis. In general, the CUSUM schemes have been shown to be very efficient in detecting small shifts in the mean of a process; they are effective even when it is desired to detect shifts in the mean that are $2\,\sigma$ or less.

 In each energy channel, we combined the CUSUM method with a novel technique based on a statistical analysis of the parameters (mean and standard deviation) used by CUSUM to estimate the uncertainties of the onset times. In the cases where the assumptions about the statistics of the sequence of random variables are not valid, namely, the random variable does not have a normal probability density function and the monitored process is not stationary, then the selection of the time interval where the background mean and the standard deviation is calculated can produce a significant uncertainty in the estimated onset times.

We estimate the uncertainty of the determined onset times using a Monte Carlo analysis of the CUSUM parameters, which are calculated inside the pre-event time interval. We separate the initial time interval into randomly distributed sub-domains of variable sizes and we calculate the signal parameters in each iteration. Then, we estimate an onset time for every iteration using the CUSUM method. If the assumptions about the signal statistics are not valid, then different signal parameters are calculated in every iteration, and therefore, the onset time estimation with the CUSUM method could be different for every parameter set. In most of the cases this process produces a collection of estimated onset times that are normally distributed around a certain fixed value. The maximum of the distribution corresponds to the proton onset time with the highest probability and its width to the onset time uncertainty.

 The velocity dispersion analysis
(VDA) of the time when the first particles arrive at the detector  is a method that is frequently used to estimate the release time of the energetic particles and their interplanetary travel path length. This method relies on the fact that the less energetic particles arrive at Earth later than the more energetic particles. The VDA is based on the assumptions that the first arriving particles of all energies are simultaneously released from the same acceleration site in the solar corona and their propagation is scatter free in the interplanetary medium. We should also add as a necessary condition to the above-mentioned assumptions that the energetic particles have an energy density lower than that of the local magnetic field to be effectively confined during their propagation in magnetic flux tubes without significant cross-field transport. In Appendix~\ref{ap:EnrgDensCal} we calculate the energy density for a selected, very energetic event to prove that the energy density of the energetic protons is negligible compared to the magnetic energy density.
  
The velocity dispersion equation at 1 AU can be written as
\begin{equation}\label{eq:VDA}
t_{rel} = t_{onset}(E) - s \cdot \beta^{-1}(E) + 8.33\,min.,\end{equation}

\noindent where $t_{onset}(E)$ is the observed onset time of the first arriving particles at 1 AU, $\mathrm{t_{rel}}$ is the release time of protons with energy E at the Sun, $s$ is the apparent path length travelled by the particles, and $ \beta^{-1}(E) = \frac{\mathrm{c}}{v(E)} $ where $v\mathrm{(E)}$ is the relativistic velocity of protons with energy E. From the linear relationship between the arrival times of the first particles measured on Earth $t_{onset}(E)$ and the $\beta^{-1}(E)$ the solar release time can be estimated from the intersection at $\beta\mathrm{(E)}^{-1}=0$ and the interplanetary distance travelled by the energetic particles is estimated from the slope.

We constructed the VDA plots from the onset times that were determined by the CUSUM method at different energies and then we plotted them versus $\beta^{-1}(E)$. We also included the uncertainty of the onset time determination in every energy channel and the uncertainty of the corresponding $\beta^{-1}(E)$ value, which is determined from the width of every energy channel. For the linear fit, we used the method of \cite{York2004} , which applies the maximum likelihood estimation approach for the determination of the slopes, the intercepts, and the standard errors of a number of data sets. This method produces a weighted least-squares fit of a straight line to a set of points with error in both coordinates.

The result of the VDA for the 15 January 2005 event is presented in the middle panel of Fig.~\ref{Fig:VDAprev}. We mark with blue crosses the onset times and their uncertainty, derived from the Poisson-CUSUM method and the Monte Carlo method, respectively. The red line corresponds to the linear fit of the onset times derived from the maximum likelihood estimation method. From Eq.~\ref{eq:VDA} we calculated the apparent path length and the proton release time at the Sun (the 8.33 minutes of travel time appear in Eq.~\ref{eq:VDA} to enable the comparison between the arrival times of protons and the electromagnetic radiation at 1 AU). In the case of Fig.~\ref{Fig:VDAprev} the release time was determined to be at 06:23 UT with an uncertainty of $\pm3.0$\,minutes and the apparent path length 1.67 AU with an uncertainty of $\pm0.08$\,AU. 

 Usually, the onset times are strongly affected by fluctuations of intensities in the individual energy channels, which may introduce a distortion and produce a large error both in the apparent path length and the release time. In our analysis, we considered that when the apparent path length values were outside the range of 1\todash{}3 AU, the VDA and the related release times should not be trusted; thus, 32 out of 115 events were rejected from our analysis. Nominally, the path length should be close to the length of the Parker spiral from the Sun to the Earth (\textit{i.e.} $\sim$1.15\,AU, for solar wind velocity $\mathrm{\upsilon_{sw}=440km/s}$). However, several velocity dispersion studies \citep[\textit{e.g.}][]{Krucker2000,Vainio2013} have shown significant deviations between the path lengths that were obtained from the VDA and the Parker spiral length. These deviations can arise, in cases of larger path lengths ($>$\,1.5\,AU), when energetic particles are subjected to strong interplanetary scattering \citep[][path lengths from simulated data]{Lintunen2004} or when the interplanetary magnetic field is disturbed (\textit{i.e} when it has a significant component out of the ecliptic plane). Moreover, deviations for path lengths $<$\,1.0\,AU can arise from instrumental effects \citep[\textit{e.g.}][in the case of energetic electrons]{Haggerty2002,Haggerty2003,Krucker2001}. Another reason is that the hypothesis used in VDA that particles at all energies are simultaneously released at the same distance from the spacecraft may be wrong.

\begin{table}[!t]
\centering \caption{\textbf{Overview of the Radiospectrographs.}}
\label{Table:RadioObserv}
\begin{tabular}{c c c c c } 
Radio           & Observing & Frequency & Temporal \\
Spectograph     & Time (UT) & (MHz) & Resolution \\
\hline \\[-1.0em]
\hline \\[-1.0em] \\[-1.0em]
Wind/Waves & All Day      & 0.02\,--\,13.825        & 1\,min            \\
\hline \\[-1.0em]
Artemis-IV & 06:00\,--\,15:00 & 20\,--\,650$^\dagger$  & 1\,sec \\
\hline \\[-1.0em]
Culgoora   & 20:00\,--\,07:00 & 18\,--\,1800 & 3\,sec   \\
\hline \\[-1.0em]
RSTN       &  All Day & 25\,--\,180     & 3\,sec        \\
\hline
\end{tabular}\\
$^\dagger$: Before July 2002 the frequency range was 110\,--\,650~MHz. \\
\end{table} 
 
\subsection{Composite radio spectra}

To compare  the proton release time with the associated radio emission for the 83 selected events directly, we assembled composite dynamic radio spectra. The high-coronal to interplanetary (IP) parts of the dynamic spectra consisted of \textit{\textup{Wind/Waves}} data \citep{Bougeret1995} with one minute resolution from 20\,kHz to 13.825\,MHz, from the two \textit{WAVES} receivers known as RAD1 (20\,kHz\,--\,1.04\,MHz) and RAD2 (1.075\,--\,13.825\,MHz). The lower-coronal parts in our composite spectra consisted of data from the following ground instruments (radio spectrographs): \textit{ARTEMIS-IV} \citep[][]{Caroubalos01,Kontogeorgos2006}, \textit{Culgoora} solar radio spectrograph \citep{Prestage1994}, or  \textit{Radio Solar Telescope Network} \citep[RSTN:\,][]{Guidice81}. We occasionally used \textit{Nancay Decameter Array} (DAM) quick-look spectra from the radio survey project\footnote{http://secchirh.obspm.fr/}. We used ARTEMIS-IV data from the sweep frequency receiver ASG with a temporal resolution of 0.1 second, which we integrated over 1 second. An overview of the radio data we used is presented in Table~\ref{Table:RadioObserv}.


\begin{figure*}    
   \centerline{ \includegraphics[width=0.80\textwidth]{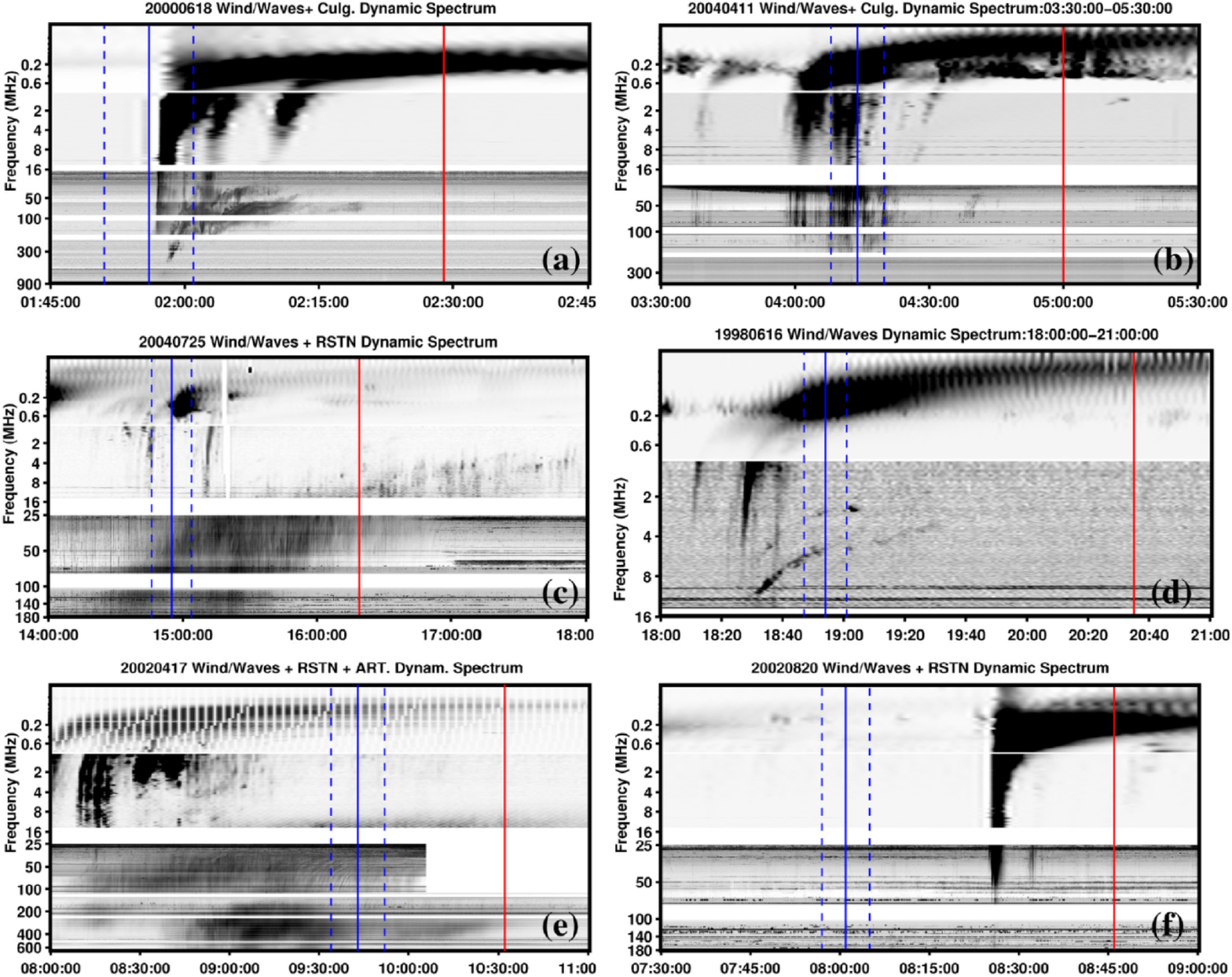} }
              
\caption{Radio spectra for different radio-associated cases: (a) event of 18 June 2000 with type III/II radio association; (b) event of 11 April 2004 with type III association; (c) event of 25 July 2004 with type III/IV-c association; (d) event of 16 June 1998 with type II association, (e) event of 17 April 2002 with type IV-c association; and (f) event of 20 August 2002 with ``no inferred'' radio association. In each panel, the vertical lines have the same meaning as in Fig.~\ref{Fig:VDAprev}.
        }
   \label{Fig:PrevSpect}
   \end{figure*}

For the event of Fig.~\ref{Fig:VDAprev}, we present the associated composite radio spectrum from Wind/Waves and ARTEMIS-IV on which we have marked the arrival (red line) of the 54.8\,--\,80.3 MeV protons at the detectors on L1. We also marked the release time (solid blue line) as inferred from the VDA analysis and its uncertainty from the linear fit (dashed blue lines). The dashed blue lines mark the ``release window''. For the 15 January 2005 event, we identified type III, II, and IV radio bursts events within the release window.

\section{Radio emission associated with SEPs}\label{sec:GenRadio}

\subsection{General remarks}

For the 83 selected events, we identified the types of radio bursts that occurred within each release window. If transient solar radio emission (Type III,\,II,\,IV-continuum) was present within the proton release window of an SEP event, we categorized it as an SEP with ``inferred radio association''. In 18 cases no transient solar radio emission was observed within the release window and these events have been characterized as SEPs with ``no inferred association'' (see example in Figure~\ref{Fig:PrevSpect}). A further examination of these events showed that significant radio emission was evident relatively near ($\pm$30 minutes) the release window in all cases. In Fig. \ref{Fig:PrevSpect}.(f) we present the composite radio spectra of a ``no inferred association'' case for the 20 August 2002 event. In this event there was a strong group of type III bursts 25 minutes after the end of the interval ascribed to the release window. We believe that the inferred lack of association is caused by failures in the VDA due to the incorrect determination of onset times from the Poisson-CUSUM method. 

\subsection{Cases with inferred radio association}\label{subsec:AssCaseRadioOcc}

After the rejection of the 18 ``no inferred radio association'' events we classified the remaining 65 cases according to the radio emission within the release window. Type III bursts were registered in 53 cases, type II bursts were registered in 44 cases and type IV-c bursts in 15 cases. In Table~\ref{Table:AllRadioAss} we give the proton release time and the apparent path length with their uncertainties, as inferred by the VDA for these 65 cases with ``inferred radio association''; we also give the radio associations for every event.

 We further separated the radio associations into more groups according to the combination of radio bursts that occurred within the release window. For example, if a type III and a type II radio burst both occurred within the release window we registered the event as ``type III/II associated'' or if only a type III occurred within the release window, we registered the event as ``type III-only associated''. In Fig. \ref{Fig:PrevSpect} we present the composite radio spectra for different cases of radio association. For the resulting radio associations we specifically found: 25 cases for III/II; 18 cases for type III-only; seven cases for type II-only, and 15 cases for the mixes of IV-c with III or II.  The association rate for type III-only events is lower than the rate of type III/II group and the association rate for type II-only cases is significantly lower than the type III-only group. A synopsis of the above radio association rates is presented in Table~\ref{Table:RadioAssosOver}.
 
We also investigated the radio emission associated with the 65 events without taking  the VDA error bars into account. In every SEP event we registered  the radio emission within $\pm1$~hour from the proton release time; this range is always larger than the VDA errors. With this selection we have a general view of the radio burst occurrence, near the proton release time, of SEPs. We found that in ten cases we have type III-only occurrences; in 34 cases type III and type II; and in 21 cases mixes of type IV-c occurrence (i.e. IV-c, III/IV-c, II/IV-c, III/II/IV-c). From the association rates, without the use of the VDA errors, we note that the type III/II occurrence is still the highest (see Table~\ref{Table:RadioAssosOver}). For the cases with only type III occurrence we note that their percentage survives, albeit smaller, even if we consider a time window of $\pm$1 hour from the release time. All the type II events are accompanied by type III radio bursts within the time window of $\pm$1 hour.
 
\begin{table}[!t]
\centering \caption{\textbf{Overview of the SEPs' Radio Associations.}}
\label{Table:RadioAssosOver}
\begin{tabular}{c c c c c c c} 
\multicolumn{6}{c}{Radio Association Cases:} \\
 & III/II & III & II & mix/IV-c & Total & No Assoc.\\
\hline \\[-1.0em]
\hline \\[-1.0em] \\[-1.0em]
$^\dagger$ & 25         & 18                            & 7                       & 15           &       65      & 18    \\
\hline \\[-1.0em]
 & 38$\%$       & 28$\%$                   & 11$\%$              & 23$\%$       &         100$\%$ &       \\
\hline \\[-1.0em]
$^\ddagger$ & 34                & 10                            & 0                       & 21           &       65      & 0     \\
\hline \\[-1.0em]
 & 52$\%$       & 16$\%$                   & 0$\%$               & 32$\%$       &         100$\%$ &       \\
\hline
\end{tabular}\\
$^\dagger$: Radio association rates \textit{\textup{within}} proton release window. \\
$^\ddagger$: Radio association rates within $\pm$1~h. from proton release time. \\
\end{table}

With the above analysis we confirm the well-established result of \cite{Cane2002} that SEPs above $\sim$\,20 MeV are accompanied by type III bursts (in our case 53/65 events; \textit{i.e.}\,80$\%$). However, from our analysis it is also evident that among these 53 type III associated events there are 33 cases that are also accompanied by a type II radio association (33/53 events; \textit{i.e.} 62$\%$). In general, type II bursts were registered in 44/65 cases (\textit{i.e.} 68$\%$).


\subsection{Flare- or shock-related particle release process? }\label{subsec:FlareShocksep}

From the separation of the radio associations into groups, we have concluded that a clear-cut distinction between the flare-related and CME-related SEP events is difficult to establish. In several cases the proton release is accompanied by both a type III and a type II radio burst, which are tracers of flare- and shock-related particle release processes, respectively.  The above conclusions apply if we assume that the radio signatures of both flare- and shock-related particle release processes inside the release window can both contribute to the proton acceleration.

In the cases of type III/II association it is hard to tell if one or both related processes can contribute to the particle release process. However, we investigated whether the association of the proton release times with type II radio bursts indicates a separate acceleration process of the SEPs. The results of our investigation appear in  Section~\ref{subsec:FlareShocksep}. 

\subsubsection{Radio spectral characteristics}
 For the type III/II associated events we further examined the SEP composite radio spectra and we determined whether the type III bursts emanated from the type II bursts. For the cases in which this is true, we could argue that the shock wave accelerates both the type III emitting electrons and the SEPs. We have considered that the type III radio burst emanates from the type II in the cases where type II start precedes or coincides with the start of the type III. We have not included in the above group the cases where the difference between the type III and type II start is less than $\sim$4 minutes (\textit{e.g.} the case in Fig.~\ref{Fig:PrevSpect}.a).

\begin{figure*}    

\centerline{\includegraphics[width=0.95\textwidth]{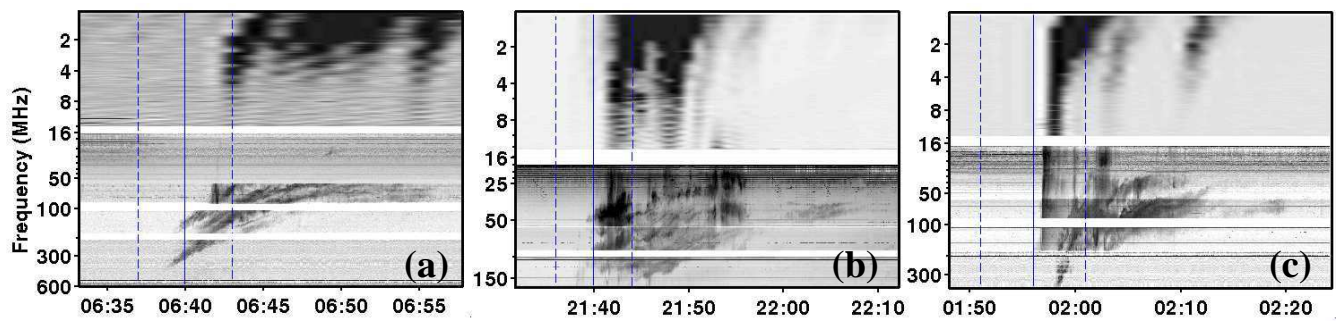}  }
     \vspace{-0.052\textwidth}   
     \centerline{\Large \bf     
      \hspace{0.300\textwidth}  \color{black}{(a)}
      \hspace{0.260\textwidth}  \color{black}{(b)}
      \hspace{0.265\textwidth}  \color{black}{(c)}
         \hfill}
 
     \vspace{0.028\textwidth}    

\caption{Radio spectra for different type III/II radio-associated cases: (a,b) events of 22 November 1998 and 19 December 2002 for which the type II radio burst precedes the start of the type III; in these cases the type III emanates from the type II; (c) event of 18 June 2000, the type III precedes the type II radio burst and we cannot argue that the type III emitting electrons are accelerated by the shock wave. In each panel, the vertical blue lines have the same meaning as in Fig~\ref{Fig:VDAprev}.}
   \label{Fig:sepEveEmanate}
   \end{figure*}

From this analysis we have found that in 18 cases the type III emanates from the type II (Group 1), while, in seven cases it does not (Group 2). Additionally, for the cases with type III/II/IV-c association we found that five events fall into the first group, while three events comprise  the second group. In the majority of the events we consider, we did not find the classical sequence of radio emission in which the type II bursts follow the type III radio emission. This result provides a good argument in support of the scenario that the type III emitting electrons are accelerated by the shock wave. In Fig.~\ref{Fig:sepEveEmanate} we present the composite radio spectra for different type III/II radio-associated cases and we show two events (Fig.~\ref{Fig:sepEveEmanate}.a-b) for which the type II radio burst precedes the start of the type III and one event (Fig.~\ref{Fig:sepEveEmanate}.c) that probably does not show this temporal pattern.

\subsubsection{Proton energy spectra}

We have also examined whether the type III/II and II-only associated SEP events have a flatter/harder energy spectrum than the type III-only events. For the 65 events with ``inferred radio association'', we have constructed their proton energy spectrum by taking the peak flux in each energy channel and presenting it as a function of the proton energy. If we ignore propagation effects then we can argue that the produced peak flux spectrum is representative of the injection spectrum \citep{Lin1982,Krucker2007}. Although, the peak times occur later for lower energies due to velocity dispersion, the peak flux spectrum would not be affected from this dispersion. Additionally, the calculated spectral slopes from the peak  flux spectrum cannot depend on the time interval where the calculation is performed.

We have not included the SOHO/ERNE
low-energy channels (1 to 10 MeV) in the proton energy spectra analysis to exclude events with streaming-limited peak intensities that could cause a ``plateau'' in the energy spectra and also avoid any contamination that may originate from solar sources that are not related to the SEP event under study. In all the cases, the proton energy spectra showed a break in the power law with a steeper (softer) component at higher energies; this break in the spectrum occurs around 25 MeV. The calculation of the spectral power-law indices has been performed by fitting two power-law functions to the observed energy spectra, one to the low- and one to the high-energy component. This break up in the spectrum can be either ascribed to a direct signature of the acceleration process or to a secondary process, such as wave-particle interactions, particle escape from the acceleration region, or transport effects.

For the derived spectral indices ($\mathrm{\gamma}$), we found values within the range -1.27 to -4.99 for the low-energy component (LEC) of the spectrum and within the range -1.67 to -4.99 for the high-energy component (HEC) of the spectrum. We separated spectral indices into groups according to the radio association of each SEP event (see Sect.~\ref{sec:GenRadio}.2). In Table~\ref{Table:SectralIndex} we present a synopsis of the mean values for $\mathrm{\gamma_{LEC}}$ and $\mathrm{\gamma_{HEC}}$ for every separate radio association group. The separation of $\mathrm{\gamma}$ into the above radio association groups showed that, in the cases with type II association (\textit{i.e.} III/II, III/II/IV-c, II/IV-c, II-only), the mean $\mathrm{\gamma}$ values are higher than the $\mathrm{\gamma}$ mean values for the cases with type III association without a type II burst (\textit{i.e.} III-only, III/IV-c). 

To validate whether the differences between the derived group means are significant, we performed an analysis of variance (ANOVA) for both the $\mathrm{\gamma_{LEC}}$ and $\mathrm{\gamma_{HEC}}$ values. The one-way ANOVA compares the means between the groups we are interested in and determines whether those means are significantly different from each other. We excluded from the analysis the cases of III/IV-c and II/IV-c association because the statistical sample was small. From the ANOVA we found that the difference of mean $\mathrm{\gamma}_{LEC}$ between the type III-only and type III/II groups is statistically significant (p-value\,$=$\,$2\times10^{-5}$). Similar results were found from the comparison of mean $\mathrm{\gamma}_{HEC}$ values but with higher p-values. Therefore, the spectra of the groups with type IIs are harder than those with type III-only bursts and this result is statistically significant. This is an indication that the shock related type II is contributing to the SEP acceleration.

Moreover, we compared the mean $\mathrm{\gamma}$ values between the groups of type III/II, II-only, III/II/IV-c association. We found that the ANOVA did not show a clear separation for those groups (p-value\,$=$\,$0.18$ for LEC and $0.08$ for HEC); all the events with type II association could belong to the same population of spectral characteristics. The type II-only group mean $\mathrm{\gamma}$ values should be closer to a shock related proton acceleration scenario without a flare related counterpart. However, the determined mean $\mathrm{\gamma}$ values for the type II-only group was found to have relatively high value and this might be explained by the fact that we failed to find any type II-only cases that are completely isolated from type III radio bursts, \textit{i.e.} cases in which the only radio burst in the vicinity of the proton release time is a type II burst. We conclude that the spectral hardening of the type II cases compared to the type III shows that the shock related type II is contributing to the SEP acceleration.

\begin{table}[!t]
\centering \caption{\textbf{Overview of the SEPs' energy spectrum indices.}}
\label{Table:SectralIndex}
\begin{tabular}{c c c c c c c} 
\multicolumn{6}{c}{Radio Association Cases:} \\
 & \small{III/II} & \small{III} & \small{II} & \small{III/II/IV-c} & \small{III/IV-c$^\dagger$} & \small{II/IV-c$^\dagger$} \\
\hline \\[-1.0em]
\hline \\[-1.0em] \\[-1.0em]
$\mathrm{\gamma_{LEC}}$ & -2.41         & -3.22         & -2.79          & -2.35         &       -4.25   & -2.28         \\
\hline \\[-1.0em]
$\mathrm{\gamma_{HEC}}$ & -2.98         & -3.54         & -3.60          & -2.84         &       -3.75   & -2.77 \\
\hline \\[-1.0em]
$\mathrm{I_{max}}^\ddagger$        & 0.043              & 0.023                 & 0.005        & 0.111            &       0.002   &       0.106   \\
\hline
\end{tabular}\\[0.40em]
$^\dagger$: Associations with small statistical sample. \\
$^\ddagger$: Maximum SEP intensity in the energy range 54.8--80.3 MeV. \\
\end{table}

\subsubsection{Maximum SEP intensity}

Furthermore, we examined the maximum SEP intensity in relation to the radio association cases. Earlier studies showed a relation between the peak SEP intensities and the associated CME properties, such as CME speed \citep[\textit{e.g.}][]{Kahler2001,Gopalswamy2004}. The correlation between SEP intensity and CME speed has been considered  strong evidence supporting the scenario that SEPs are accelerated by CME-driven shocks. We expect that the type II associated SEP events would be more intense than the events without a type II association if we assume that the type II-related shock is contributing to the SEP acceleration. However, the scatter of the maximum SEP intensity could vary over 3 to 4 orders of magnitude; any correlation with maximum SEP intensity would have a high uncertainty.

For events with ``inferred radio association'', we determined the maximum proton intensity ($I_{max}$) in the energy range from 54.8 to 80.3 MeV. We found that the events with type III/II association exhibit higher values of $I_{max}$ compared to the events with type III-only association. The group with type III/II/IV-c association has the highest value of $I_{max}$ , meanwhile the type III/IV-c showed the lowest value. In Table~\ref{Table:SectralIndex} we present the mean values of $I_{max}$ for every radio association group. We further tested the statistical significance of the $I_{max}$ mean value differences between the radio association cases. The ANOVA showed that the differences of $I_{max}$ mean values for all the cases are not statistically significant; therefore, we cannot conclude whether the differences originate from the presence (or absence) of a shock related particle release process; \textit{i.e.} for the type III/II and the type III-only group we found a very high p-value ($p=0.45$). The large scatter of $I_{max}$ values, which probably originate from particle transport effects, certainly affect the significance of the $I_{max}$ difference between the groups.


\subsection{Acceleration in post-CME current sheet}

 We further examined the composite radio spectra  for the 15 type IV-c associated events (see Sect.~\ref{sec:GenRadio}), and investigated whether moving type IV bursts were present. For the cases that moving type IV bursts could be detected, we could argue that the energetic particles could have been accelerated at the reconnecting current sheet behind the CME \citep[see ][]{Klein2014}.

From our analysis we found seven cases (out of the 15 cases with type IV-c association) with moving type IV bursts. For these events, the particle acceleration could have taken place at the post-CME current sheet. The most characteristic cases, associated with moving type IV bursts, are the events of 25 July 2004 (Fig.~\ref{Fig:PrevSpect}.c) and 17 April 2002 (Fig.~\ref{Fig:PrevSpect}.e). In those events the moving type IV burst consisted of a prolonged radio emission with prominent drift from centimetric to decametric waves. In summary, we  found that in almost half of the cases with type IV-c association the particles could have been accelerated at the post-CME current sheet. In six out of the seven cases with moving type IV bursts, the relative contribution of particle acceleration at the post-CME current sheet is difficult to evaluate because the moving type IV burst coexists with other transient radio features inside the release window.

\section{Locations of the SEP-related flares}\label{sec:locations}

Statistical analysis of SEPs \citep[\textit{e.g.}][]{Cane1988,Cliver2004,Kurt2004,Gopalswamy2008} have shown that the solar activity associated with SEPs most often occurs in the western hemisphere. From the distributions of the ``source longitude'' of the associated flares \citet{Reames1999} and \citet{Nitta2006}  concluded that source regions of the impulsive SEP events are distributed about the longitude of best magnetic connection to the observer. The small spread in the longitude distribution could be attributed to changes in connection longitude resulting from variations in solar wind speed. A large scatter in longitude away from the nominal Parker spiral connection has been observed for the gradual events. \citet{Reames1999} has shown that the distribution of the associated flares longitudes for gradual events is nearly uniform across the solar disk. Furthermore, \cite{Klein2008} showed that there are cases where open flux tubes diverge rapidly with height and these flux tubes connect the parent active region to the Parker spiral at the source surface even when the active region is as far as 50$^\circ$ away from the nominal connection longitude. 

For the SEP events with ``inferred radio association',' we investigated the locations of the associated flares \citep[see Table 3 of][which is based on \cite{Cane2010}]{Vainio2013}. In most cases (45/65) the associated flares were located in the western hemisphere while there was a significant lack of events in the eastern hemisphere (5 events only). In 15 cases we did not find an associated flare apparently because the events originated from far behind the west limb. Both groups with type III-only association and type III/II had six back-sided cases each; this means that there are cases where the radio emission at higher frequencies could be occulted by the limb.

\begin{figure}[!t]
\begin{center}
\includegraphics[width=0.45\textwidth]{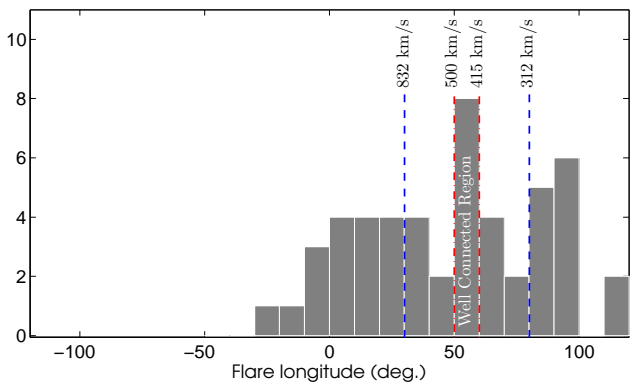}
\vspace{-0.015\textwidth}   
\end{center}
   \caption{Distribution of the SEP associated flare longitude in the cases with inferred radio association. The red dash-dotted and blue dashed lines correspond to different well-connected magnetic flux tubes, according to Parker's model of the interplanetary magnetic field, with their longitudes calculated from Eq.~\ref{equ:longSEP} using different solar wind velocities ($\upsilon_\mathrm{sw}$). The red lines correspond to $50^\circ$\,--\,$60^\circ$  western longitude ($\upsilon_\mathrm{sw}$\,$\sim$\,$415-500$~km/s). The right blue dashed line corresponds to longitude $\phi_\mathrm{0}$\,$=$\,$80^\circ$ ($\sim$\,$312$~km/s) and the left blue dashed line corresponds to $\phi_\mathrm{0}$\,$=$\,$30^\circ$ ($\sim$\,$832$~km/s). We also label the $50^\circ$\,--\,$60^\circ$ longitudes as ``Well-Connected Region''.}
         \label{Fig:FlarePos}
   \end{figure}

In Fig.~\ref{Fig:FlarePos} we present the distribution of the associated flare longitudes. The maximum of the distribution was found in the well-connected region of $50^\circ$\,--\,$60^\circ$ western longitude, but there is a significant scatter on either side of this region from $-30^\circ$ to $120^\circ$. If we accept Parker’s model for evolution of the interplanetary magnetic field and a solar wind velocity $\upsilon_\mathrm{sw}$\,$\sim$\,450~km/s, then the interplanetary magnetic flux tubes have solar connection longitudes near $50^\circ-60^\circ$ western longitude. The equation of the solar connection longitude to the connection region (\textit{i.e.} Earth) is given by the following equation \citep[\textit{e.g.}][]{Nolte1973}:

\begin{equation}\label{equ:longSEP}
\phi_0 = \frac{\omega r}{\upsilon_\mathrm{sw}}(1-\frac{R_S}{r})
.\end{equation}

\noindent Here $r$ is the distance from the Sun centre to the connection point (in our case, $212~\mathrm{R_\odot}$ for the Sun-Earth L1 point), $\omega$ is the angular speed of solar rotation (\textit{i.e.} the equatorial solar rotation period of $24.47$~days or $\omega$\,=\,$2.97\times10^{-6}$~rad/sec), $R_S$\,=\,$2.5~\mathrm{R_\odot}$ the heliocentric distance beyond which the magnetic field is assumed to take on the form of the Parker spiral, and $\upsilon_\mathrm{sw}$ is the radial solar wind speed, assumed constant in space. In Fig.~\ref{Fig:FlarePos}, we label the well-connected region of $50^\circ$\,--\,$60^\circ$ western longitude with the two red lines, calculated from Eq.~\ref{equ:longSEP} for solar wind velocity, $\upsilon_\mathrm{sw}$\,$\sim$\,$415$\,--\,$500$~km/s; the blue lines correspond to $\upsilon_\mathrm{sw}$\,$\sim$\,$832$~km/s ($\phi_\mathrm{0}$\,$\sim$\,$30^\circ$) and $\upsilon_\mathrm{sw}$\,$\sim$\,$312$~km/s ($\phi_\mathrm{0}$\,$\sim$\,$80^\circ$). 

We further separated the SEPs into groups, according to the radio associations inferred from the analysis of Sect.~\ref{subsec:AssCaseRadioOcc}. We separately analysed  the cases in which we had type III association without type II (Group A); the cases with type II association without a type III (Group B) or the mixed cases with type III and type II association (Group C). We found that, for the first group, the associated flare longitudes were only located in the western hemisphere  between longitudes  $30^\circ$ to $80^\circ$ , for 13 out of 14 cases. For the second group, we found that in four out of eight cases the flare occurred within the range $-15^\circ$ to $30^\circ$, in two cases at longitudes from $30^\circ$ to $80^\circ$, and in two cases at longitudes from $80^\circ$ to $120^\circ$. Lastly, the analysis of the third group showed a more mixed situation; in nine cases, the flare occurred inside the region of $30^\circ$ to $80^\circ$ western longitude; in another nine cases, at longitudes from $80^\circ$ to $120^\circ$; and, in six cases, at longitudes from $0^\circ$ to $30^\circ$. Also, three cases were located in the eastern hemisphere. The flare locations of the type III-related cases are more concentrated in the well-connected region, whereas the flare locations of the type II-related cases are more scattered from this region.


\section{Proton and electron release timing}\label{sec:Temporal}

\subsection{Proton and type III timing}\label{subsect:protTiii}


For the SEPs with inferred radio association, we extended our study to the timings between the radio emission of type III and the proton release time. Considering that type III radio emission is a secondary product of Langmuir waves produced by electron beams \citep[\textit{e.g.}][]{Lin1981,Lin1986}, we used type III bursts as proxies of the initial electron acceleration and escape. From the composite spectra we determined the start time of the associated type IIIs (\textit{i.e.} inside the release window). We defined the start of a radio burst type III as the point where the radio emission was above the background level by three standard deviations, preferably at the highest frequency, given that the type III would extend to the interplanetary space. In Col.~8 of Table~\ref{Table:AllRadioAss}, we present the start time of the associated type IIIs.

\begin{figure}[!t]
  \begin{center}
    \centerline{ \includegraphics[width=0.45\textwidth]{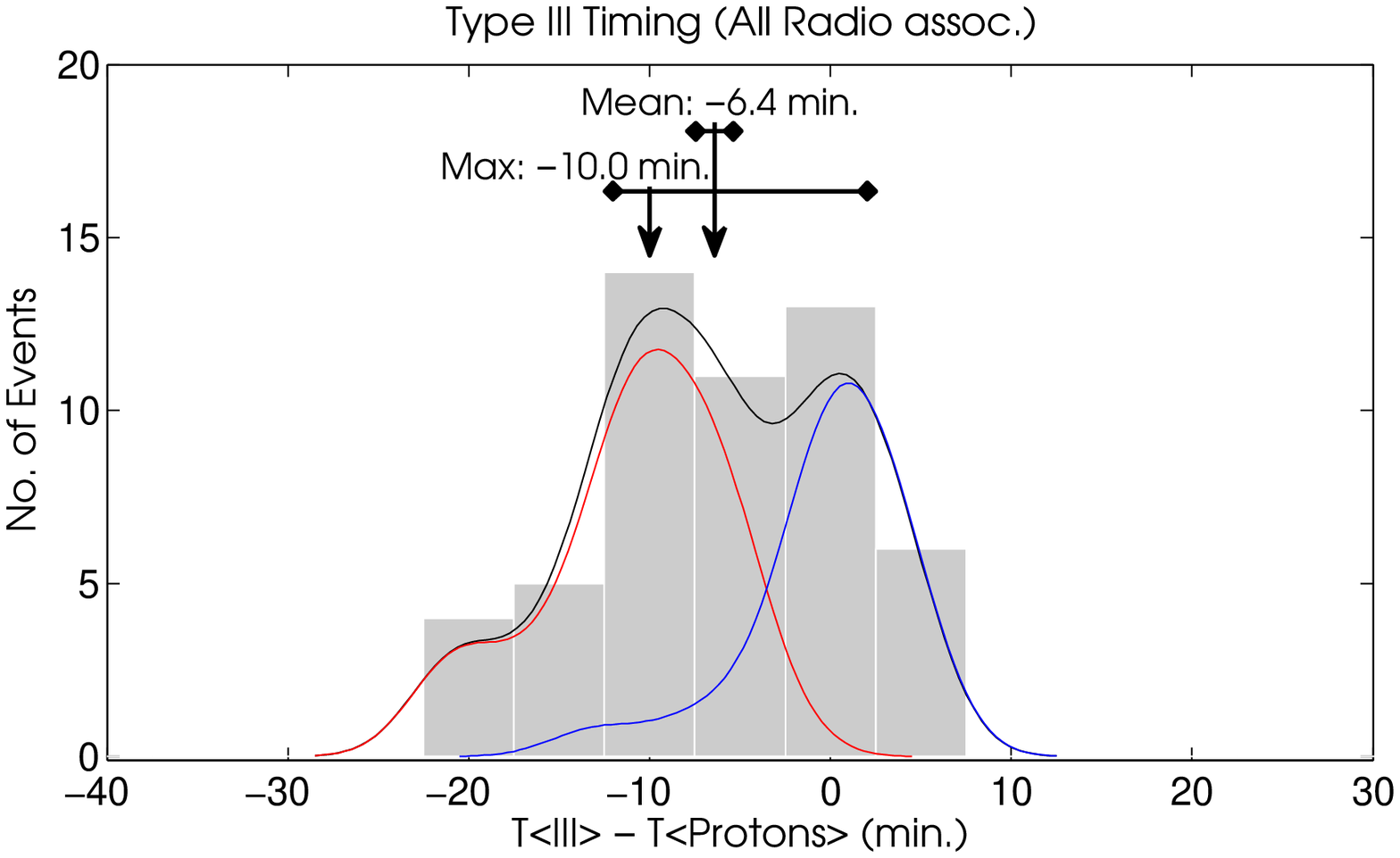} }
     \vspace{-0.25\textwidth}   
     \centerline{\bf     
      \hspace{0.065\textwidth}  \color{black}{(a)}
         \hfill}
     \vspace{0.225\textwidth}    
%
    
    \centerline{ \includegraphics[width=0.45\textwidth]{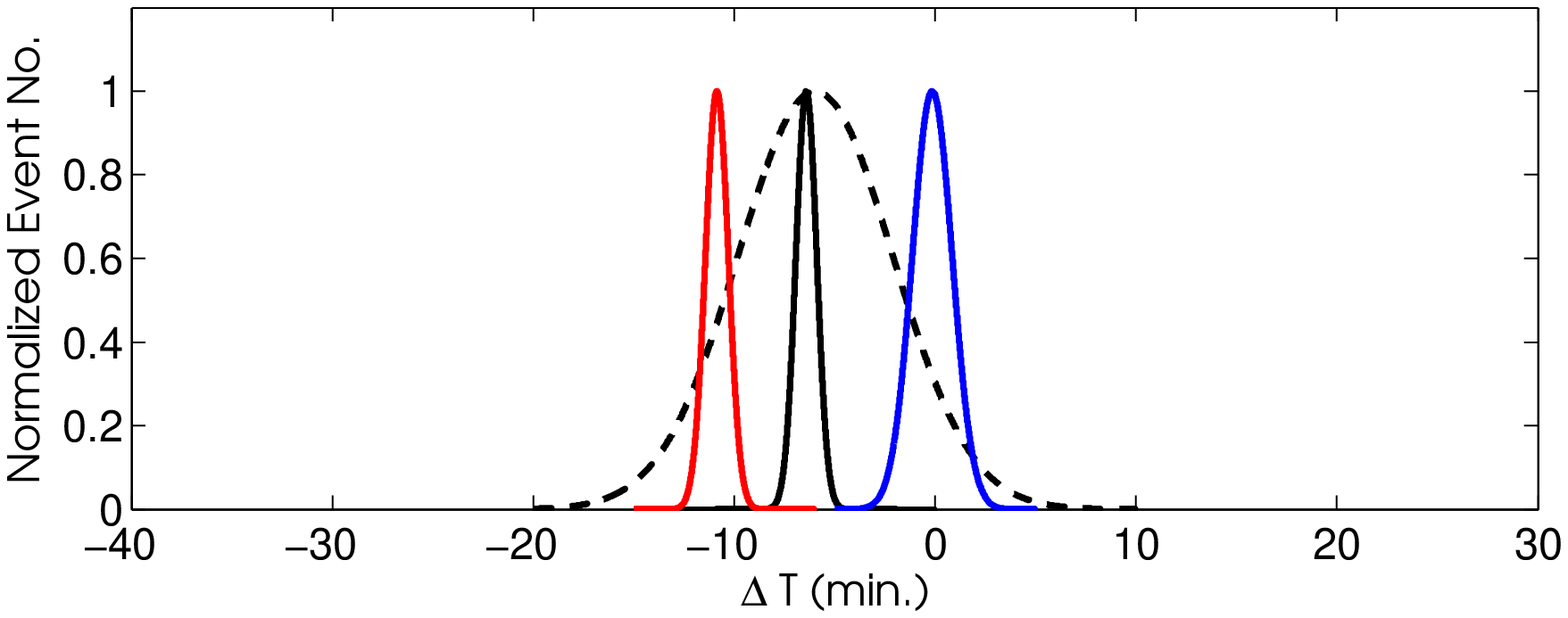} }
      \vspace{-0.17\textwidth}   
    \centerline{\bf     
      \hspace{0.065\textwidth}  \color{black}{(b)}
         \hfill}
     \vspace{0.003\textwidth}   
    \centerline{\bf     
      \hspace{0.230\textwidth}  \color{black}{}
         \hfill}
    \centerline{\bf     
      \hspace{0.300\textwidth}  \color{black}{}
         \hfill}
     \vspace{0.065\textwidth}    
  \end{center}
     \caption{(a) The distribution of the time difference between type III start and the proton release time derived from the VDA. With the arrows we denote the mean and maximum value of the distribution and with the horizontal lines we mark the errors as inferred from the Monte Carlo analysis. The solid lines show the distribution functions of the type III-proton release timings using all the events (black) and the separate cases of delayed (red) and simultaneous events (blue). (b) The best-fit Gaussians of the type III-proton release timings mean and maximum values (black solid and dashed lines, respectively) are derived from the Monte Carlo analysis using all the events. The red and blue solid lines show the best-fit Gaussians to the distributions for the mean values of the simultaneous and delayed cases, respectively. }
      \label{Fig:TIIItimInj}
\end{figure}

We found that in all but 12 events, the type III-related electron release occurred before or at the time of proton release (time differences from 0 to -24 minutes). The timing distribution (Fig.~\ref{Fig:TIIItimInj}a) showed a maximum at $-$10 minutes with the electron release preceding the time of proton release and a mean value at $-$6.4 minutes.  We further separated the events according to their exact radio associations and examined their timing distributions with proton release time. The best synchronization was exhibited by events that showed type III/II emissions, with a mean delay $-$5.3 minutes. For the events with only type III association we found a mean delay $-$5.6 minutes between type III and proton release. The worst synchronization was exhibited by events that showed an association with type IV-c radio bursts.

To validate the results from the proton-type III timing histogram, we implemented a separate analysis taking  the computed release time uncertainty $\delta$t into account. We follow the method presented by \cite{Krucker1999} to distinguish between simultaneously released events within the statistical uncertainties when $(t_{III}-t_{rel}) / \delta t$\,$>$\,$-1$, and delayed events when $(t_{III}-t_{rel}) / \delta t$\,$\leq$\,$-1$. From the above separation, it follows that when the proton release occurred before the type III start, these events could only be characterized as simultaneously released; in the case where $(t_{III}-t_{rel}) / \delta t$\,$>$\,$1$ according to the radio association analysis (see Section~\ref{sec:GenRadio}) these events have been characterized as SEPs with ``no inferred association". For 31 cases (59$\%$), the type III-related electron release occurred well before the time of proton release, and therefore, these events were delayed; for 22 cases (41$\%$) the type III-related electron and proton release was  simultaneous. In Fig.~\ref{Fig:TIIItimInj}a,  the solid black line denotes the smoothed distribution function of the time difference between type III start and the proton release time using all the events, and  the blue and red solid lines denote the distribution function for the separate cases of simultaneous or delayed events. The smoothing of the histograms was estimated with a kernel smoothing function that returns a probability density estimate for a given number of samples. We have scaled the probability density estimates for illustration purposes. The timing distribution of the delayed events (red line) showed a maximum at $-$9.5 minutes with the type III start preceding the time of proton release and a mean value at $-$10.9 minutes. Additionally, for the simultaneous events (blue line), the timing distribution showed a maximum at $-$1 minute and mean value at $-$0.14 minutes.

As a consistency check of our timings and  to estimate the uncertainty of the presented mean values, we performed a Monte Carlo analysis. With this method we verified if the above results are independent of the selection and the width of the release window. If we assume that the timing distribution is not sensitive to time shifts of the release time, within the error limits of the VDA, then if we randomly perturb the release time, the distribution characteristics should not change. In our Monte Carlo analysis, the perturbation of the measurements was performed by adding  the release error multiplied by a random number between $-$1 and 1 to every proton release time. With this method, we produced randomly distributed perturbed release times within the release error. We repeated the above procedure several times ($>10^4$~times) to have a statistically significant sample. To analyse the statistical sample produced by the Monte Carlo analysis, we tried two different approaches. At first, for every ``perturbed'' release sample, we determined the mean value of the time differences and, from the derived mean values, we analysed their distribution. Additionally, we applied the above described method using the distributions maximum values for every ``perturbed'' release sample.

In Fig.~\ref{Fig:TIIItimInj}.b we present the results of the Monte Carlo analysis for all the events (black solid and dashed lines for the mean and maximum values, respectively) and the separate cases (blue and red solid lines for the mean values of the simultaneous and delayed cases, respectively). The distributions of the mean values (solid lines, Fig.~\ref{Fig:TIIItimInj}.b) are localized near the centre line of the timing distribution functions and their maximum coincides, as expected, with the mean values of the initial timing distributions (Fig.~\ref{Fig:TIIItimInj}.a). This indicates that any randomly distributed time shifts of the initial proton release times would produce timing samples that would have mean values between the limits of the Gaussian-fitted functions for each case (black, red, and blue fitted lines, Fig.~\ref{Fig:TIIItimInj}.b). With the second (maximum) method, we found that the distribution is more scattered than that derived from the first (mean) method and this effect is caused by the samples' discrete form and the histogram binning. In fact, the maximum value can significantly change between the iterations (``perturbations'') and compared with the mean value, which is more localized and  negligibly affected between the iterations, this second approach is less reliable.
 
We used the results of the Monte Carlo analysis to match the uncertainty of the time differences' mean and maximum values. In Fig.~\ref{Fig:TIIItimInj}.a we present the mean and maximum values with their error bars for the timing distribution of all the events. The size of errors for both mean and maximum in Fig.~\ref{Fig:TIIItimInj}.a are the $\sim$\,$\pm3 \sigma$ values of the Monte Carlo analysis distributions (confidence value: $99.73\%$). For the separate cases, we found that the simultaneously released events have a mean time difference 0.1$\pm$5.0 minutes and the delayed events have a mean time difference -10.9$\pm$3.5 minutes.
 
\subsection{Proton and Wind/3DP electrons timing} \label{sec:ProtElWindTim}

\begin{figure}[!t]
\begin{center}
 \centerline{ \includegraphics[width=0.45\textwidth]{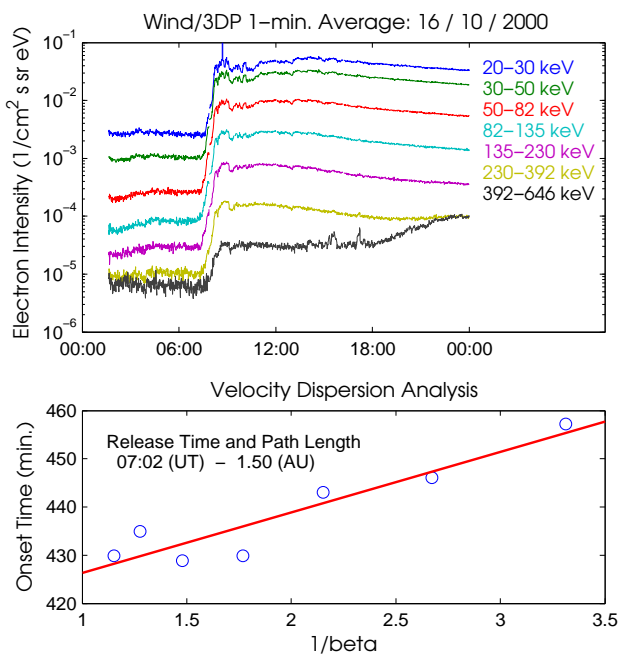}  } 
 \vspace{-0.045\textwidth}
\end{center}
   \caption{Electron event of 16 October 2000. 
   Top: electron intensity as recorded from Wind/3DP for 7 electron energy channels between 20 keV and 646 keV. 
   Bottom: electron velocity dispersion analysis. We mark  onset times with blue circles and  linear fit with a red line. }
         \label{Fig:PreviewElectron}
   \end{figure}

We also analysed the electron data from the Wind/3DP for all the type III associated events. In all (53) of these proton events we found electron events. We identified the onset times for seven energy channels between 20 keV and 646 keV and we performed velocity dispersion analysis using all the available energy channels. 

For the VDA, we followed the same procedure as described in Sect.~\ref{Subsect:ProtRelTime} with the exception that we did not use the Poisson-CUSUM method for the onset determination. The Wind/3DP backgrounds are generally not steady and often show non-dispersive variations on various timescales, which  can affect any statistical method of the onset determination. We determined the Wind/3DP onset times by eye, which, compared to the statistical onset determinations (\textit{i.e.} Poisson-CUSUM), has the advantage of using the background and rise phase profile information on both short and long timescales \citep{Kahler2006,Kahler2007,Haggerty2002}.

We performed the VDA and, in most cases (34/53), we found a reasonable linear fit. We rejected the events in which no dispersion between the Wind/3DP energy channels was observed (15/53) or the path length values were outside the range 1-3~AU (4/53). A comparison between the electron's and proton's path lengths showed that protons travel considerably longer path lengths (s$_{p}=2.15$\,AU) compared to the electrons (s$_{e}=1.51$\,AU). The longer path length for protons implies that protons are subjected to more scattering from the Sun to the spacecraft than the electrons. \cite{Malandraki2012} also found and confirm a similar discrepancy between the path lengths of protons and electrons using VDA for a case study.

In Fig.~\ref{Fig:PreviewElectron} we present an example for the 16 October 2000 event. The top panel shows the electron intensity in seven energy channels. We have added  $8.33$~minutes to the derived electron release time to enable the comparison between the arrival times of electrons and the electromagnetic radiation at 1 AU. For this event, the dispersion between the different energies is pronounced. From the electron VDA (Fig.~\ref{Fig:PreviewElectron}, bottom), we calculated the electron release time at 07:02~UT with an uncertainty of $\pm5$~minutes, and we found an apparent path length of $1.50$~AU with an uncertainty of $\pm0.29$~AU.

\begin{figure}[!t]
  \begin{center}
        \centerline{\includegraphics[width=0.45\textwidth]{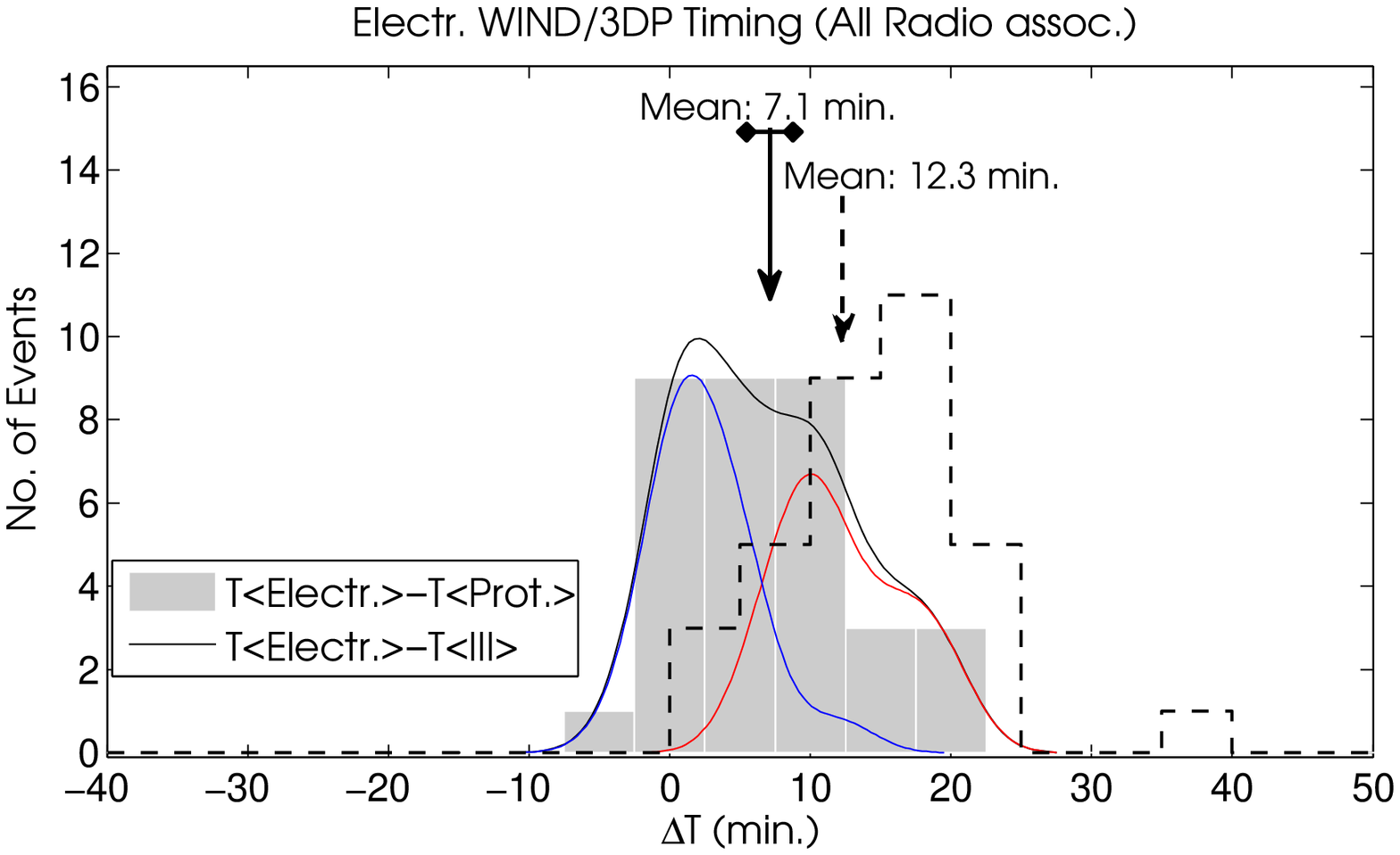}}
          \vspace{-0.25\textwidth}   
     \centerline{\bf     
      \hspace{0.065\textwidth}  \color{black}{(a)}
         \hfill}
     \vspace{0.225\textwidth}    
     
        \centerline{\includegraphics[width=0.45\textwidth]{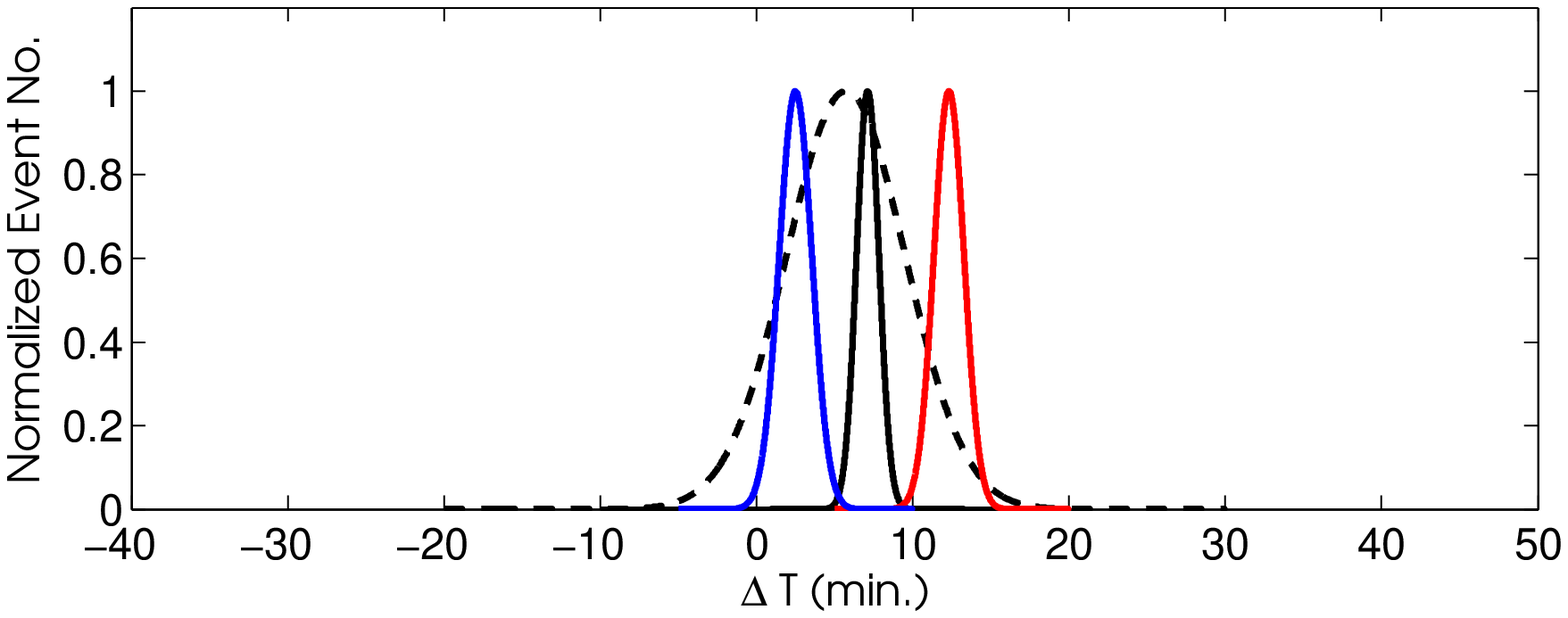}}
     \vspace{-0.175\textwidth}    
        \centerline{\bf     
      \hspace{0.065\textwidth}  \color{black}{(b)}
         \hfill}
     \vspace{0.115\textwidth}   
  \end{center}
     \caption{(a) The grey histogram shows the distribution of the time difference between electron release derived from the VDA of Wind/3DP electron data and the proton release time, derived from the VDA of SOHO/ERNE proton data. The dashed histogram shows the distribution of the time difference between electron release derived from the VDA of Wind/3DP electron data and the type III start. With the solid line arrow and the dashed line arrow we mark the mean values of the electron-proton and the electron-type III timing difference, respectively. The solid lines show the distribution functions of the type III-proton release timings using all the events (black) and the separate cases of delayed (red) and simultaneous events (blue). (b) The best-fit Gaussians of the electron-proton release timings mean and maximum values (black solid and dashed lines) derived from the Monte Carlo analysis using all the events. The red and blue solid lines show the best-fit Gaussians to the distributions for the mean values of the simultaneous and delayed cases, respectively.}
     \label{Fig:WindeTimInj}
\end{figure}

To compare the proton release from SOHO/ERNE and the electron release from the Wind/3DP we plotted the distribution of their time difference (Fig.~\ref{Fig:WindeTimInj}.a). From the characteristics of the distribution, we note that there is a time delay of $\sim$7 minutes of the electron release times with respect to the proton release times. We found that the electron release with the use of the Wind/3DP data occurred in most of the cases (33/34, $\sim$97$\%$) after or at the time of the proton release. Specifically, we found a mean value of $7.1$~minutes and a maximum at $5.0$~minutes (between -2.5 and 12.5) for the electron-proton release time difference distribution (see the solid arrow in Fig.~\ref{Fig:WindeTimInj}.a). We further separated the electron-proton release time differences into simultaneously released events and delayed events following the method presented in Sect~\ref{subsect:protTiii} for the separation of the proton-type-III time differences. The uncertainty of the relative timings would be, in this case, the sum of the proton/electron release uncertainties in quadrature, $\delta t$ =$ \sqrt{\delta t_e^2+\delta t_p^2}$, where $\delta t_e^2$ and $\delta t_p^2$ is the electron and proton release uncertainty, respectively. From the separation of the timings into simultaneous and delayed events we found that in 18 ($\sim$53$\%$) events the proton and electron release process is simultaneous, while for 16 ($\sim$47$\%$) events the electron release is delayed compared to the proton release.

To validate whether the above time differences is independent of both proton and electron VDA error we performed Monte Carlo analysis in the same way as described in Sect.~\ref{subsect:protTiii}. In this analysis we only changed the bin size of the maximum distribution because the statistical sample was small. We present the results of the Monte Carlo analysis in Fig.~\ref{Fig:WindeTimInj}.b., where it is clear that the above timing results are independent of the proton-electron timing errors induced by the VDA fit. 

To better illustrate the relative time difference of Wind/3DP electron and type III release, we selected to over-plot in Fig.~\ref{Fig:WindeTimInj}.a, the relative timing between the electron release from Wind/3DP and the start of the type III radio burst, presented with dashed line. We note that the electron release occurs on average $\sim$12.3 minutes after the start of the type III. For the events in which the Wind/3DP electron release occurred after the start of the proton-associated type IIIs (33/34), we found that in all but three cases, the durations of the type IIIs (mean $\sim$4 minutes) were smaller than the electron-type III start time differences. If we additionally consider the uncertainty of the electron release then only six events could be characterized as simultaneously released with the type III (\textit{i.e.} 28/34 were delayed, 82\%). Therefore, the Wind/3DP electron release typically occurs well after the end of the proton-associated type IIIs.

\begin{figure}[!t]
\begin{center}
\includegraphics[width=0.47\textwidth]{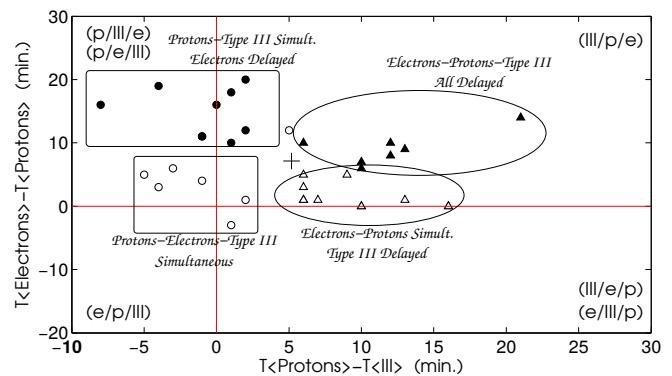}
\vspace{-0.015\textwidth}   
\end{center}
   \caption{Scatter plot of the proton-type-III time difference versus the electron-proton time difference. We use circles for the proton-type-III simultaneously released events and triangles for the delayed events. For the proton-electron release, we use unfilled markers for the simultaneously released events, and black-filled markers for the delayed events. The small black cross represents the point of the mean value for each time difference. We used the red horizontal and vertical lines to separate the scatter plot into categories according to the labelled cases (\textit{i.e.} III/p/e, III/e/p, e/III/p, p/III/e, p/e/III, e/p/III). We label the four resulting groups, as derived from the separation, into simultaneously released and delayed events; we use rectangles and ellipses only for illustration purposes. }
         \label{Fig:SumReltime}
   \end{figure}

To further investigate other possible implications that can be derived by the proton-electron relative timings, we present in Fig.~\ref{Fig:SumReltime} the proton-type-III time difference versus the electron-proton time difference. With this representation we can separate the cases into groups according to the sequence of the type III-proton-electron occurrence. For example, the events in which the type III occurs first, followed by the proton and the high-energy electrons release are located in the labelled region ``(III/p/e)''. Most of the events (19/34) are located in this region, which means that type III proton and electron release most probably occurs in this successive order. However, there are also some cases that do not fit with the III/p/e occurrence scenario, for example, there are cases in which the proton release is delayed compared to the electron release (III/e/p) or the proton release occurs before the type III (p/III/e). An example where electrons are detected prior to protons has been given by \citet{Malandraki2012}. For p/III/e cases (seven events), we note that all the events have relative time differences between the type III and proton release, which are lower than the proton release uncertainty; the same applies to the single case of the III/e/p group for the relative time differences between the electron and proton release. For the remaining combinations, we found one case in p/e/III group.

Taking the $\delta t$ uncertainties into account, we  also considered the separation of the events into simultaneously released within the statistical uncertainty and delayed events as discussed in Sect.~\ref{sec:ProtElWindTim}. In Fig.~\ref{Fig:SumReltime} we represent the proton-type-III simultaneously released events with circles and the delayed events with triangles; for the separate cases of the proton-electron release we use unfilled markers for the simultaneously released events and black-filled markers for the delayed events. We label in Fig.~\ref{Fig:SumReltime} the four groups as derived from the above separation. Comparing the number of events in each group we found small differences; the highest number of events (10) was found in the group where the protons and electrons are simultaneously released and both are delayed compared to the release of the type III emitting electrons. We note that when we add the electron's release in our analysis and taking  the $\delta t$ uncertainties into
account, the separation into groups according to the sequence of the type III-proton-electron occurrence becomes ambiguous. 

Summarizing the results of the relative time differences of type III bursts start, proton release from SOHO/ERNE and electron release from Wind/3DP (high-energy electrons) we have found that the proton, the high-energy electron release and the type III start do not show a dominant release sequence. We found a simultaneous proton release compared to the high-energy electron release (53$\%$); additionally we found that both the proton (59$\%$) and the high-energy electron (82$\%$) release were delayed compared to the type III start. However, not all the events fit with the above scenario, which means that the release sequence of protons, the high-energy electrons, and the type III emitting electrons is more complex than a single release scenario.

\subsection{Electron release associated radio emission}

\begin{figure}[!t]
\begin{center}
 \includegraphics[width=0.45\textwidth]{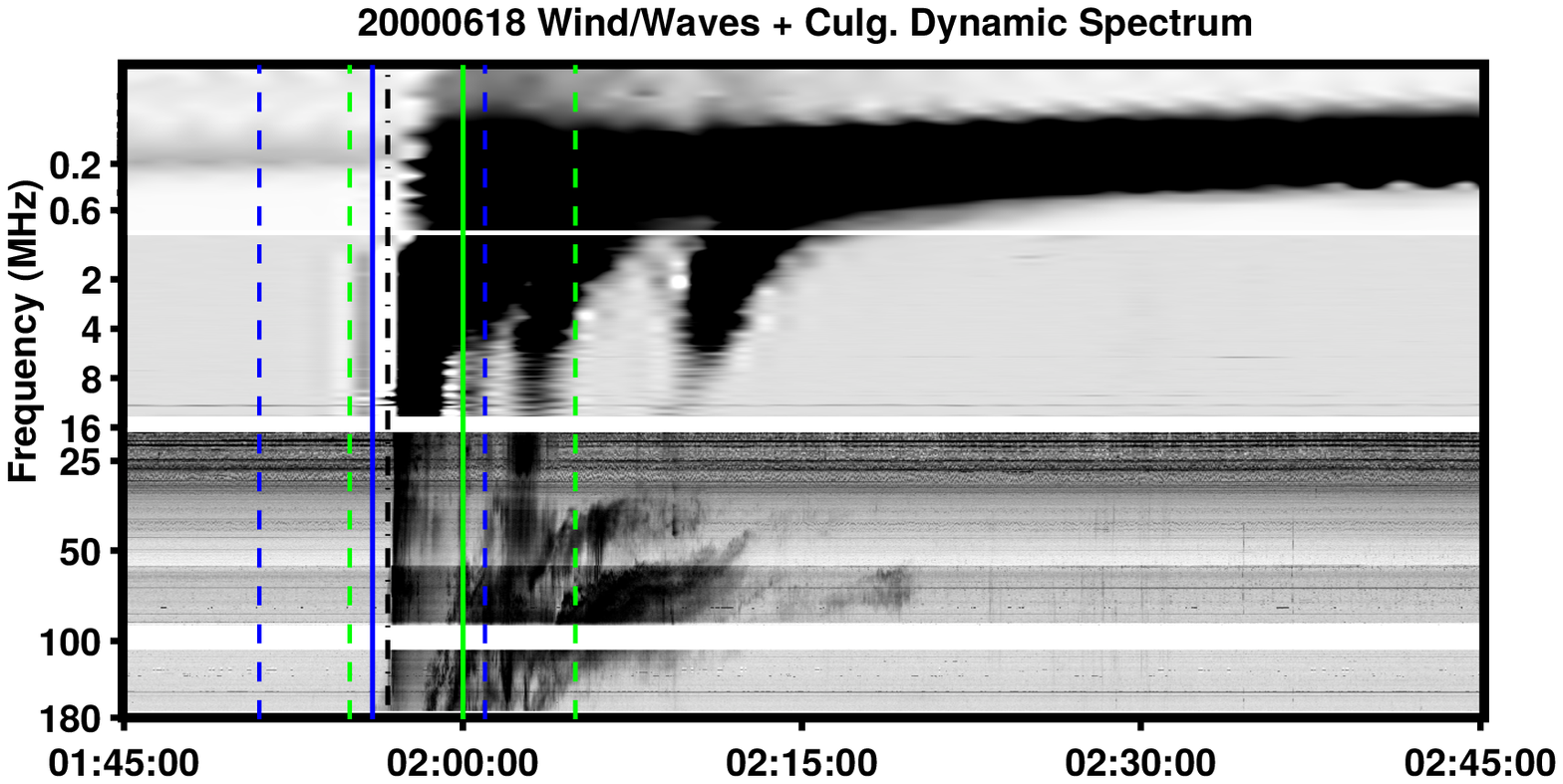}
 \includegraphics[width=0.45\textwidth]{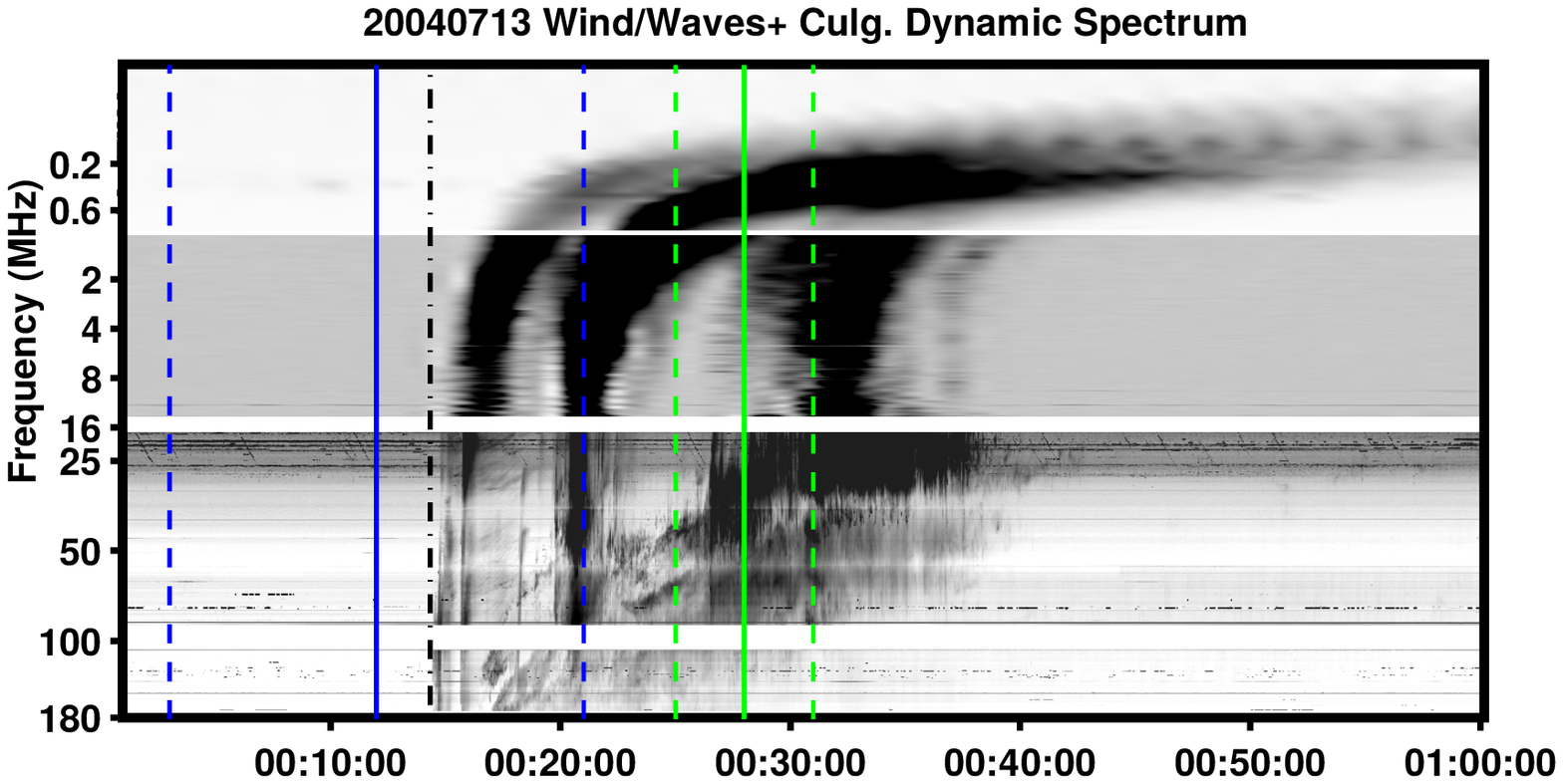} 
\vspace{-0.015\textwidth} 
\end{center}
   \caption{Composite radio spectra for two different electron events. Top: event of 18 June 2000. Bottom: event of 13 July 2004. With the vertical blue line and dashed lines we present the proton release time derived from the VDA analysis and its uncertainty, respectively; with the vertical green line and dashed lines we present the electron release time derived from the VDA analysis and its uncertainty, respectively; with the black dash-dotted line, we present the start of the type III burst.}
         \label{Fig:PreviewElectronSpectra}
   \end{figure}

For the 34 SEP events with an electron event observed by Wind/3DP and for which we found a reasonable VDA fit, we investigated the associated radio emissions with the electron release time as inferred by the electron VDA. To identify the types of radio bursts that occurred within each electron release window, we over-plotted the electron release time and its uncertainty from the least-squares fit at the composite radio spectra (see Fig.~\ref{Fig:PreviewElectronSpectra}, for an example). We registered the associated radio bursts for every event following the same procedure as described in Sect.~\ref{sec:GenRadio}.  In Fig.~\ref{Fig:PreviewElectronSpectra} we present the composite radio spectra for two cases: (1) the event of 18 June 2000 where the electron release time (green lines) is simultaneous compared to the proton release (blue lines) and the type III (black line) burst, and (2) the event of 13 July 2004 where the electron release time lags compared to the proton release and the type III burst. 

From the inspection of the composite spectra, we registered radio bursts associated with the electron release for all the cases except from the event of 07 October 1997. For this event a possible reason for the lack of association might be that the relevant radio emission was occulted as the associated flare was located well behind the solar limb. In the remaining 33 cases with inferred radio association we registered 23 cases with type III and II association (III/II); seven cases with only a type III radio burst, and three cases with mixed type IV-c association.

\section{Estimate of proton release heights}\label{sec:Spacial}

For the 44 SEP events associated with type II bursts, we estimated the proton release heights using the assumption that type II-related shock wave actually accelerates the first particles. Since we have shown in Sect.~\ref{subsec:FlareShocksep} the accelerative role of the type II-related shock waves, the above assumption could not be rejected entirely although we cannot determine the exact relative contribution of the shock-related and flare-related processes to the SEP acceleration. For each event, we determined the corresponding CME observed by the SOHO's \textit{Large Angle and Spectrometric Coronagraph} \citep[LASCO: ][]{Brueckner1995} and registered on the online catalogue of \cite{Yashiro2004}. We found associated CMEs for all the events, except one case during a SOHO/LASCO data gap. In all cases there was only one CME near the onset of the proton event and specifically near the proton release window. 

\begin{figure}[!t]
\begin{center}
\includegraphics[width=0.45\textwidth]{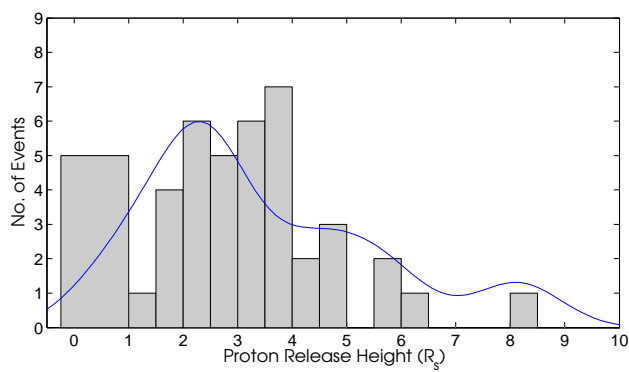}
\vspace{-0.020\textwidth}   
\end{center}
   \caption{The distribution of the estimated proton release heights from the CME leading edge height-time measurements. The blue solid line represents the distribution function of the estimated proton release heights for the SEP events whose source region longitude was greater than 80$^\circ$.}
         \label{Fig:TIIHeightDist}
   \end{figure}

For the estimate of the proton release heights, we used the information for the height of the CME leading edge as measured by the automated procedure described by \cite{Yashiro2004}. We extrapolated (or interpolated) their projected height-time measurements to the time of the proton release as estimated from the VDA. From the resulting distance, we have a value for the proton release height if we accept that the protons are primarily released at the leading edge of the CME. Recent studies have shown the importance of the lateral CME expansion as a driver of the particles into well-connected regions \citep[][]{Rouillard2012}. Another important issue is the standoff distance \citep[\textit{e.g.}][]{Ontiveros2009,Gopalswamy2011} between the CME leading edge and the shock front that might introduce an error to our estimates. Actually, the particles are not necessarily accelerated at the CME leading edge, but they can be located anywhere in the sheath region between the CME front and the shock front. Thus, our method might underestimate the ``real'' proton release height, depending on the standoff distance and the particles spatial distribution into the sheath region.

In Fig.~\ref{Fig:TIIHeightDist} we present the distribution of the resulting SEP projected release heights. The height is measured from the solar centre, so the photosphere is at a height equal to one. In eight cases, we found SEP release heights that correspond to heights below the photosphere ($<$1$\mathrm{R_\odot}$) and these cases are artefacts of the analysis. This is probably an effect of the initial assumption that the CME leading edge propagation can be approximated as a linear fit of height or result from projection effects since the trajectory of a CME is determined in projection on the plane of the sky. Our assumption that the height increases linearly with time and usually agrees with CME height-time data above 1.5\,$\mathrm{R_\odot}$, but the lack of measurements for heights below the LASCO/C2 field of view can, in some cases, affect the accuracy of the computation. 

The SEP projected release height distribution (Fig.~\ref{Fig:TIIHeightDist}) peaks at 3.5\,$\mathrm{R_\odot}$ (between 3\,--\,4\,$\mathrm{R_\odot}$) and extends up to 8\,$\mathrm{R_\odot}$; we found a mean value of 3.4\,$\mathrm{R_\odot}$, excluding from the calculation the five cases with heights below the photosphere. For the cases with type II-only association, we found for the proton projected release heights a mean value of $\sim$4.4\,$\mathrm{R_\odot}$. For these cases, we do not present the peak of the distribution because the number of cases is too small to support a statistically significant maximum. Moreover, we over-plot in Fig.~\ref{Fig:TIIHeightDist} the distribution function of the estimated SEP projected release heights only for the SEP events whose source region longitude was greater than 80$^\circ$. We used this criterion to reduce the sample of events with uncertain height-time profiles due to projection effects. From the resulting distribution we found a maximum at $\sim$2.25\,$\mathrm{R_\odot}$ and a mean value at $\sim$3.79\,$\mathrm{R_\odot}$. Additionally, the cases with unrealistic heights ($<$1$\mathrm{R_\odot}$) were significantly reduced. Our findings can be interpreted by the following scenario. As a CME expands radially outward from the solar corona, a shock wave forms when the Alfv\'{e}n speed ahead of the CME falls below the CME speed. So it is not surprising that the SEP-associated shocks accelerate particles at starting heights near 2.5\,$\mathrm{R_\odot}$ if we accept the Alfv\'{e}n speed-height dependence. According to \cite{Mann2003}, the Alfv\'{e}n speed decreases rapidly with radial distance to a minimum at $\sim$1.2\,--\,1.8\,$\mathrm{R_\odot}$, then rises to a maximum at $\sim$3.8\,$\mathrm{R_\odot}$ and decreases thereafter. 

Similar results of release heights have been reported in previous studies \citep{Huttunen2005,Gopalswamy2012,Reames2009} for proton/helium SEPs or ground-level events (GLEs). \cite{Reames2009} has studied the onset times in large SEPs, which have been detected by neutron monitors at ground level, and converted this release time to a radial distance of the source shock wave from the Sun. They found that the acceleration for well-connected events begins at 2\,--\,4\,$\mathrm{R_\odot}$. Almost the same result was found by \cite{Gopalswamy2012} for the release of several GLEs; they showed that the release occurs when the CMEs reach an average height of $\sim$3.09\,$\mathrm{R_\odot}$ (between 1.71 to 4.01\,$\mathrm{R_\odot}$). \cite{Huttunen2005} also determined the apparent proton release heights for 13 proton/helium events with the most reliable release times (path length: 1.0\,--\,1.5 AU) and they found a release height from 2 to 10\,$\mathrm{R_\odot}$. The above proton release heights have been converted to distance from Sun's centre, when necessary, to be comparable with our measurements.

\section{Summary of results}\label{sec:DiCo}

We have studied the properties of major SEP events, especially focussing on (1) the association of the SEP release time, as inferred by the velocity dispersion analysis (VDA), with transient solar radio emissions recorded by space and/or ground based radio spectrographs; and  (2) the time difference of proton release with respect to the escape of keV electrons into space, derived from both DH type III bursts and VDA. Our key findings are:

\begin{itemize}

\item{ Both flare- and shock-related particle release processes are observed in high-energy proton events at $>$50 MeV. A clear-cut distinction between flare-related and CME-related SEP events is difficult to establish.}

\item{Proton release is most frequently accompanied by both type III and II radio bursts (38$\%$), but there is a significant percentage of cases with only type III occurrence (28$\%$).}

\item{Typically, the protons are released after the start of the associated type III bursts and simultaneously or before the release of energetic electrons.}

\item{The locations of the major SEP-related flares was concentrated in the western hemisphere with a peak in the well-connected region of $50^\circ$\,--\,$60^\circ$ western longitudes and significant scatter in the range from $-30^\circ$ to $120^\circ$.}

\item{ The proton release for the type II associated cases typically occur at heights from 2.0 to 3.5\,$\mathrm{R_\odot}$.}

\end{itemize}

\section{Discussion}\label{sec:DiCo2} 

Unlike previous statistical studies, we managed to directly associate the proton release time, as inferred by the proton VDA, with different types of transient radio emission, from the composite radio spectra, within the proton release window. Based on our radio association analysis, we fail to find support for the hypothesis that only the flare-related particle release process dominates in SEP events with energy above $\sim$50\,MeV. If the latter hypothesis were true we would expect that the proton release should  mainly be associated with the occurrence of type IIIs only. However, a significant number of cases shows radio association characteristics, which resemble those of the classical group of impulsive cases where the SEPs observed in high energies are only associated with flare-related phenomena, such as the presence of type III radio bursts. Additionally, we found that, less frequently,  SEP events are only associated with shock-related phenomena, such as the presence of type II radio bursts. For the type IV-c radio associated events, we found that an acceleration of particles in the post-CME current sheet could apply in almost half of the cases, although the relative contributions from various sources is difficult to evaluate because the moving type IV bursts coexist with other transient radio features. In summary, the results of the radio association analysis leads us to conclude that, in many cases, it is possible that the flare-related and shock hypothesis of particle release could both apply. 

The above conclusions apply if we assume that all radio-related processes (flare or shock related) inside the release window can both contribute to the proton acceleration and release. In general it is hard to derive the relative contributions of these processes to the particle release process, but we investigated whether the association of the proton-release time with type II radio bursts indicates a separate acceleration process of the SEPs. From the inspection of the type III/II radio spectra, we found that in most of the cases (73$\%$) the type III bursts emanate from the type II bursts. This result supports the scenario that the type III emitting electrons are accelerated by the shock wave for those cases. As an additional argument, we computed the proton energy spectra for all the SEP events and found that the events with type II association have a harder energy spectrum than those without a type II. Summarizing the above results, we have found that the presence of a type II may indicate a contributing acceleration process from the associated shock wave.

In Sect.~\ref{sec:Temporal} we presented the relative timing between the proton release and the start of the associated type III radio emission. We used type III bursts as proxies of the initial electron acceleration and escape. In general, we found a late release for protons. In $59\%$ of our cases, the type III-related electron release occurred well before the time of proton release; these cases have been characterized as delayed. The apparent delay between the proton release and the type III radio emission can be ascribed to several reasons, such as selective acceleration, transport effects, particle diffusion, and particle trapping. An other possibility was suggested by \citet{Klein2005}, who attributed the delay of protons to the successive injection of particles in different flux tubes, which are not all connected to the spacecraft. Lastly, we did not find a dependence between the timings and the flare longitude, nevertheless, some events outside the well-connected region exhibit the worst synchronization.

 We further investigated the relative timing between the proton and electron release using Wind/3DP data. When we implemented  the $\delta t$  uncertainties in the analysis and we separated the events into simultaneously released and delayed events, we found that in $53\%$ of our cases the electron release (observed by Wind/3DP) compared to the proton release (observed by SOHO/ERNE) was simultaneous within the statistical uncertainty. As far as the remaining delayed cases are concerned we note that similar results have been presented by \citet{Cliver1982}. Of course, the comparison of our relative timings with those of \citet{Cliver1982} might be indirect mainly because of the different electron-proton energies analysed in each study, but the underlying physical reason could be similar. This delay could be attributed to turbulence or waves associated with the CME-driven shock that traps the low-rigidity electrons much more effectively than the protons, selective acceleration, or transport effects. 
 
As far as the delayed electron release is concerned, compared with the start of the type III ($\sim$12 minutes) we additionally found that in all except six cases the durations of the type IIIs (mean: 4 minutes) were smaller than the electron-Type III start time differences. Therefore, the Wind/3DP electron release typically occurs well after the end of the proton-associated type IIIs. Previous studies have investigated this delay providing different scenarios. \citet{Krucker1999} attributed this delay to two distinct populations of electrons, the low-energy population associated with the type III radio burst and the high-energy population representing the delayed electrons. Some of our analysed cases (17/33) could fit into the above scenario because the inspection of the composite spectra showed a very good association of the electron release with group of type IIIs or isolated type III separate from the initial type III, which occurs at the start of the SEP event. 

\citet{Haggerty2002} showed that the electron delay with respect to the type IIIs could be due to the acceleration and release of the near-relativistic electrons by an outgoing coronal shock. However, we did not find any dependence of the relative timings with the existence (or not) of a type II association. \citet{Cane2003} 
showed that the low- and high-energy electrons belong to the same population and, thus, interaction effects in the interplanetary medium might cause the delays. Even though the electron propagation and the interaction effects in the interplanetary medium could be a possible cause for the observed electron delays, we did not find any correlation between the relative timings and the electron apparent path lengths.

In Sect.~\ref{sec:locations} we examined the locations of the related flares for the SEPs with ``inferred radio association''. The maximum of the flare helio-longitude distribution was found in the well-connected region of $50^\circ$ to $60^\circ$ western longitude, but there is also significant scatter on either side of this region from $-30^\circ$ to $120^\circ$. We further separated the SEPs into groups, according to the radio associations inferred from the analysis of Sect.~\ref{subsec:AssCaseRadioOcc}. From this separation, it is evident that the type III-only cases are associated with flare locations that are more concentrated to the well-connected region, whereas for the cases where the proton release is associated with type IIs only (shock-related cases), the SEP-related flares could be located far from the well-connected region. A similar result has been presented by \citet{Reames1999} for the cases of the SEP associated flare locations for the impulsive and gradual cases.

For the SEP events related to type II bursts and CMEs (44 events), we estimated the proton release heights from the height-time profile of the CME leading edge at the moment of the proton release as inferred from the VDA. The height distribution of proton release peaks at 2.5\,$\mathrm{R_\odot}$ (between 2\,--\,3\,$\mathrm{R_\odot}$) and extends up to 8\,$\mathrm{R_\odot}$. Previous studies have presented similar results for the release heights of SEPs and GLEs \citep{Huttunen2005,Gopalswamy2008,Reames2009}. Unlike \citet[][]{Reames2009}, we found no dependence of the SEP release height with the associated flare longitude. 

\begin{acknowledgements}
A.K., A.N., and E.V acknowledge support from the SEPServer project, funded from the European Commission’s Seventh Framework Programme. A.K., A.N. acknowledge support by European Union (European Social Fund -ESF) and Greek national funds through the Operational Program Education and Lifelong Learning" of the National Strategic Reference Framework (NSRF) -Research Funding Program: "Thales. Investing in knowledge society through the European Social Fund.” The authors would like to thank Karl-Ludwig Klein, Rami Vainio, Wolfgang Dr{\"o}ge, Neus Agueda, Athanasios Papaioannou, and Urs Ganse for useful discussions that helped to improve the manuscript, and the referee Karl-Ludwig Klein for his valuable comments and suggestions that led to  significant improvement of the quality of the manuscript.
\end{acknowledgements}

\bibliographystyle{aa} 
\bibliography{SepBiblio}

\appendix

\section{Calculation of energy density of energetic protons}\label{ap:EnrgDensCal}

Here we investigate the contribution of the energetic particles to the total energy content of the plasma by calculating the energy density of a major SEP event at 1\,AU and by comparing it with the energy density of the local magnetic field. In the case where the computed energy density of the energetic protons ($\epsilon_\mathrm{p}$) at 1 AU exceeds the local magnetic field energy density ($\epsilon_\mathrm{B}$), the particles are no longer confined into magnetic flux tubes and they can modulate the local magnetic field. The proton energy density can be derived from the following equations:

\begin{equation}\label{equ:PenergDensi}
\epsilon_\mathrm{p} = \int_\Omega^{\,} \int_{E_1}^{E_2} \frac{E}{\upsilon}\,\frac{\mathrm{d}N}{\mathrm{d}E} \, \mathrm{d}E \, \mathrm{d}\Omega = \int_\Omega^{\,} \int_{E_1}^{E_2} \frac{E^2}{\beta\,c}\,\frac{\mathrm{d}N}{\mathrm{d}E} \, \mathrm{d\,ln}E \, \mathrm{d}\Omega
,\end{equation}

\noindent where $E$ is the particle energy, $\Omega$ is the instrument solid angle, and $\beta$ is the particle velocity in $\upsilon/\mathrm{c}$. The integration over the angle $\mathrm{\Omega}$ is calculated for the particle detector effective angle (field of view) and the integration over the energy is calculated from the minimum $(E_\mathrm{1})$ to the maximum ($E_\mathrm{2}$) energy of the detected protons. The magnetic energy density is given by the following equation:

\begin{equation}\label{equ:Benerg2}
\epsilon_\mathrm{B} = \frac{B^2}{2\,\mu_0}~~~~[J/m^3]
.\end{equation}

\noindent In the above equation, the magnitude of the magnetic field is measured in Tesla and the magnetic energy density is in $\mathrm{Joules/m^3}$.

\begin{figure}
\begin{center}
\includegraphics[width=0.40\textwidth]{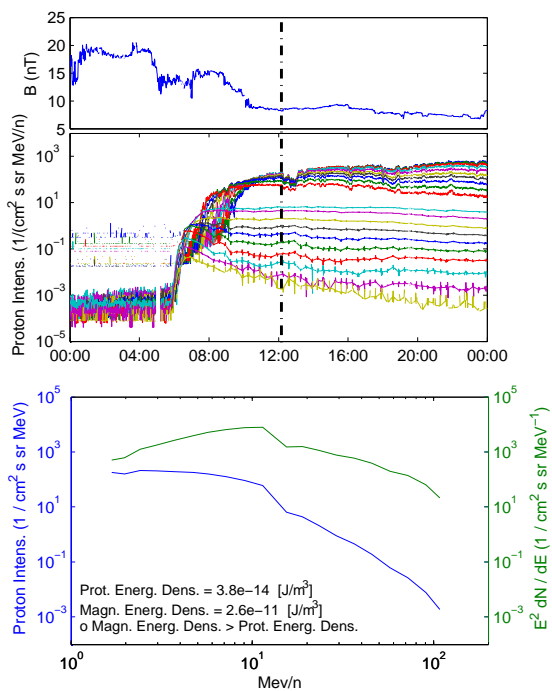}
\vspace{-0.025\textwidth}   
\end{center}
   \caption{Data for the calculation of the proton and magnetic energy density for the SEP event of 26 December 2001 at 12:10\,UT. Top: average of the magnetic field vector from Wind/IMF. Middle: proton intensity in 20 energy channels of SOHO/ERNE. The vertical black dash-dotted line marks the time of the energy density calculation. Bottom: proton energy spectrum in logarithmic scale at 12:10\,UT shown with a blue line and the measure of $E^2\mathrm{d}N/\mathrm{d}E$ shown with a green line. }
         \label{Fig:EnergDenc}
   \end{figure}

We selected to analyse the SEP event of 26 December 2001. To derive proton spectra, we used the SOHO/ERNE particle data of one minute average values. The SOHO/ERNEs' particle detectors field of view \citep[half angle;][]{Torsti1995} is $\sim$30$^\circ$ for the low-energy detector (LED) and $\sim$60$^\circ$ for the high-energy detector (HED). Combining the data of the HED and the LED, we have an energy coverage from 1.68 to 108 MeV. We first estimated the SEP energy spectra and then we integrated $E\,\mathrm{d}N/\mathrm{d}E$ to derive the $\epsilon_\mathrm{p}$. For the calculation of the energy spectra, we chose not to perform a time shift between the different energy channels because we wanted to compare $\epsilon_\mathrm{p}$ with the measurements of the local magnetic field, at 1\,AU and not at the acceleration site. For the computation of the $\epsilon_\mathrm{B}$, we used one minute average data from Wind \textit{Magnetic Field Investigation} (MFI) instrument, for the magnitude of the average magnetic field vector.

In Fig.~\ref{Fig:EnergDenc} we present the proton intensity and the magnitude of the magnetic field for the event of 26 December 2001. At 12:10\,UT we calculated the proton energy spectra (blue line) and the measure of $E^2\,\mathrm{d}N/\mathrm{d}E$ (green line). The latter in logarithmic scale is proportional to the integral in Eq.~\ref{equ:PenergDensi} and its value shows the energy region where the contribution to the integral is maximum; in our case the maximum contribution to the proton energy density comes from protons with energy close to 10\,MeV. The magnetic energy density was  $\epsilon_\mathrm{B}$=2.6\,10$^{-11}$\,$\mathrm{J/m^3}$ and this value is almost three orders of magnitude larger than the calculated proton energy density whose value was $\epsilon_\mathrm{p}$=3.8\,10$^{-14}$\,$\mathrm{J/m^3}$.

From the above analysis, it is evident that the magnetic energy density for the 26 December 2001 event is much larger than the non-thermal proton energy density. We repeated the same analysis for the very energetic events ($\mathrm{I_p>0.2}$~pfu) of 26 October and 02 November of 2003 and we found similar results; the lowest difference between $\epsilon_\mathrm{B}$ and $\epsilon_\mathrm{p}$ was almost two orders of magnitude.

\section{Table of events with inferred radio association}\label{ap:TableAll}

In Table~\ref{Table:AllRadioAss} we present our results for the 65 proton events with ``inferred radio association''. In columns 2 and 3 we give the proton release times and path lengths, respectively, with their uncertainties as inferred from the VDA. Also, in column 4 we give for every event the resulting transient radio emission associated with the proton release according to the analysis of Sect.~\ref{sec:GenRadio}. In Col.\,5 we present the locations of their associated flares that were used in the analysis of Sect.~\ref{sec:locations}. The "--" entries correspond to flares occurred far behind the west limb. The electron release times and path lengths with their uncertainties as inferred by the VDA of Sect.~\ref{sec:ProtElWindTim} are given in columns 6 and 7, respectively. The estimated start times of the type III bursts used in the relative	 timings of Sect.~\ref{subsect:protTiii} are presented in Col.\,8.

\begin{table*}
\centering \caption{\textbf{Events With ``Inferred Radio Association''.}}
\label{Table:AllRadioAss}
\begin{tabular}{c c c c c c c c c|}
\hline
\hline  \\[-0.5em]
 Event Day& Proton		& Proton Path  	& Proton Rel.	   & Flare	& Electron	& Electron Path		&  Type-III   \\
 	  & Rel. (UT)		& Length (AU)	& Radio Assoc.	   & Long.	& Rel. (UT)	& Length (AU)	&  Start (UT)   \\
\hline  \\[-0.5em]              
                 
19971007  & 12:53\,$\pm$\,07 &  2.66\,$\pm$\,0.13   &    III         &  --   & 13:13\,$\pm$\,03	 & 1.48\,$\pm$\,0.18	& 12:51	\\
19971104  & 06:06\,$\pm$\,06 &  1.86\,$\pm$\,0.15   &    III/II/[IV-c] &  33W  & 06:13\,$\pm$\,03	 & 1.17\,$\pm$\,0.15	& 05:56	\\
19971113  & 21:30\,$\pm$\,06 &  2.49\,$\pm$\,0.12   &    III       &  --   & 21:36\,$\pm$\,02	 & 1.02\,$\pm$\,0.10	& 21:20	\\
19971114  & 13:16\,$\pm$\,10 &  2.42\,$\pm$\,0.27   &    III       &  --   & 13:16\,$\pm$\,04	 & 1.54\,$\pm$\,0.24	& 13:00	\\
19980420  & 10:15\,$\pm$\,05 &  2.67\,$\pm$\,0.12   &    III/II    &  90W  & 10:25\,$\pm$\,01	 & 1.15\,$\pm$\,0.07	& 10:03	\\
19980502  & 13:45\,$\pm$\,02 &  1.34\,$\pm$\,0.06   &    III/II/[IV-c] &  15W  &	--		 & 	-- 		& 13:34	\\
19980509  & 03:34\,$\pm$\,06 &  2.55\,$\pm$\,0.12   &    III/II/[IV-c] &  100W &      --		 & 	-- 		& 03:22	\\
19980616  & 18:54\,$\pm$\,07 &  1.80\,$\pm$\,0.11   &    II        &  115W &      --		 & 	-- 		& --	\\
19981122  & 06:40\,$\pm$\,03 &  1.71\,$\pm$\,0.08   &    III/II    &  82W  & 06:51\,$\pm$\,06	 & 1.45\,$\pm$\,0.28	& 06:41	\\
19990424  & 13:21\,$\pm$\,06 &  2.14\,$\pm$\,0.12   &    II        &  --   &	--		 & 	-- 		& --	\\
19990509  & 18:08\,$\pm$\,05 &  1.73\,$\pm$\,0.13   &    III       &  95W  & 18:13\,$\pm$\,03	 & 1.03\,$\pm$\,0.15	& 17:59	\\
19990527  & 10:41\,$\pm$\,03 &  1.77\,$\pm$\,0.08   &    III       &  --   &	-- 		 & 	-- 		& 10:38	\\
19990601  & 18:56\,$\pm$\,06 &  2.71\,$\pm$\,0.15   &    III/II    &  --   & 19:04\,$\pm$\,03	 & 1.85\,$\pm$\,0.16	& 18:44 \\
19990611  & 00:41\,$\pm$\,03 &  1.69\,$\pm$\,0.08   &    III/II    &  --   &    --		 & 	-- 		& 00:41 \\
20000212  & 04:28\,$\pm$\,13 &  2.17\,$\pm$\,0.42   &    II        &  24W  &    --		 & 	-- 		& --	\\
20000217  & 20:50\,$\pm$\,05 &  1.54\,$\pm$\,0.08   &    III/II    &  07E  &	  --		 & 	-- 		& 20:29	\\
20000218  & 09:32\,$\pm$\,05 &  1.47\,$\pm$\,0.14   &    III/II    &  --   &	  --		 &	  --		& 09:23	\\
20000303  & 02:13\,$\pm$\,06 &  2.03\,$\pm$\,0.16   &    III/II    &  60W  & 02:23\,$\pm$\,02	 & 1.02\,$\pm$\,0.10	& 02:12	\\
20000404  & 15:28\,$\pm$\,03 &  1.51\,$\pm$\,0.07   &    III       &  66W  &	--		 & 	-- 		& 15:17 \\
20000618  & 01:56\,$\pm$\,05 &  1.77\,$\pm$\,0.24   &    III/II    &  85W  & 02:00\,$\pm$\,05	 & 1.49\,$\pm$\,0.27	& 01:57	\\
20000722  & 11:26\,$\pm$\,04 &  1.84\,$\pm$\,0.09   &    II        &  56W  &	--		 &	--		& --	\\
20000912  & 12:07\,$\pm$\,05 &  2.46\,$\pm$\,0.14   &    III/II/[IV-c] &  09W  & 12:21\,$\pm$\,04	 & 1.38\,$\pm$\,0.21	& 11:46	\\
20001016  & 07:02\,$\pm$\,05 &  1.63\,$\pm$\,0.10   &    III/II    &  95W  & 07:02\,$\pm$\,05	 & 1.50\,$\pm$\,0.29	& 06:52	\\
20001025  & 10:11\,$\pm$\,06 &  2.59\,$\pm$\,0.13   &    III/II    &  120W & 10:11\,$\pm$\,05	 & 2.69\,$\pm$\,0.26	& 10:01 \\
20001124  & 05:04\,$\pm$\,05 &  2.35\,$\pm$\,0.11   &    III/II    &  03W  & 05:07\,$\pm$\,03	 & 2.15\,$\pm$\,0.20	& 04:58 \\
20010105  & 17:39\,$\pm$\,06 &  2.39\,$\pm$\,0.12   &    III       &  --   & 17:44\,$\pm$\,04	 & 2.24\,$\pm$\,0.21	& 17:33	\\
20010329  & 09:56\,$\pm$\,07 &  2.97\,$\pm$\,0.15   &    III	   &  12W  & 10:02\,$\pm$\,08	 & 1.54\,$\pm$\,0.49	& 09:59	\\
20010402  & 11:28\,$\pm$\,16 &  2.59\,$\pm$\,0.53   &    II/[IV-c]    &  62W  &	--		 &	--		& --	\\
20010410  & 05:34\,$\pm$\,05 &  2.31\,$\pm$\,0.11   &    III       &  09W  &	--		 &	--		& 05:13	\\
20010415  & 13:47\,$\pm$\,14 &  1.61\,$\pm$\,0.16   &    III/II    &  84W  &	--		 &	--		& 13:39	\\
20010507  & 12:37\,$\pm$\,03 &  1.62\,$\pm$\,0.08   &    II        &  --   &	--		 &	--		& --	\\
20010520  & 06:02\,$\pm$\,06 &  2.31\,$\pm$\,0.13   &    III/II    &  91W  & 06:18\,$\pm$\,01	 & 1.50\,$\pm$\,0.86	& 06:02	\\
20010604  & 16:29\,$\pm$\,05 &  1.72\,$\pm$\,0.09   &    III/II    &  60W  & 16:30\,$\pm$\,03	 & 1.07\,$\pm$\,0.19	& 16:22	\\
20010615  & 15:40\,$\pm$\,05 &  1.74\,$\pm$\,0.14   &    III/II    &  --   & 15:41\,$\pm$\,02	 & 1.49\,$\pm$\,0.13	& 15:34	\\
20010619  & 03:42\,$\pm$\,07 &  2.21\,$\pm$\,0.20   &    III/II    &  --   &   --		 &	--		& 03:23	\\
20010915  & 11:48\,$\pm$\,04 &  1.51\,$\pm$\,0.08   &    III/II    &  53W  & 11:46\,$\pm$\,03	 & 1.02\,$\pm$\,0.13	& 11:48	\\
20010924  & 10:17\,$\pm$\,07 &  2.63\,$\pm$\,0.16   &    III       &  23E  &	--		 &	--		& 10:18	\\
20011019  & 01:20\,$\pm$\,09 &  2.35\,$\pm$\,0.32   &    II/[IV-c]     &  18W  &    --		 &	--		& --	\\
20011019  & 16:31\,$\pm$\,08 &  2.55\,$\pm$\,0.20   &    III       &  29W  & 16:49\,$\pm$\,03	 & 1.75\,$\pm$\,0.14	& 16:30	\\
20011022  & 15:14\,$\pm$\,05 &  1.88\,$\pm$\,0.09   &    III/II    &  18E  &	--		 &	--		& 15:02	\\
20011226  & 05:27\,$\pm$\,03 &  1.37\,$\pm$\,0.07   &    III/II/[IV-c]  &  54W  &    --		 &	--		& 05:13	\\
20020127  & 13:03\,$\pm$\,06 &  2.18\,$\pm$\,0.17   &    II        &  --   &	--		 &	--		& --	\\
20020220  & 05:54\,$\pm$\,08 &  1.36\,$\pm$\,0.11   &    III       &  72W  &	--		 &	--		& 05:55	\\
20020318  & 03:18\,$\pm$\,14 &  2.98\,$\pm$\,0.44   &    II/[IV-c]      &  100W &	--		 &	--		& --	\\
20020417  & 09:27\,$\pm$\,06 &  1.84\,$\pm$\,0.12   &    [IV-c]        &  34W  &    --		 &	--		& --	\\
20020421  & 01:15\,$\pm$\,06 &  1.24\,$\pm$\,0.11   &    III/II    &  84W  & 01:20\,$\pm$\,03	 & 1.37\,$\pm$\,0.17	& 01:20	\\
20020707  & 11:30\,$\pm$\,06 &  1.87\,$\pm$\,0.10   &    III       &  95W  & 11:39\,$\pm$\,01	 & 1.00\,$\pm$\,0.07	& 11:17	\\
20020814  & 01:57\,$\pm$\,04 &  1.36\,$\pm$\,0.08   &    III/II    &  54W  & -- 		 &	--		& 02:01	\\
20020818  & 21:38\,$\pm$\,07 &  1.46\,$\pm$\,0.25   &    II        &  19W  & -- 		 &	--		& --  \\
20020822  & 01:46\,$\pm$\,06 &  2.03\,$\pm$\,0.16   &    III       &  62W  & 02:05\,$\pm$\,06	 & 1.88\,$\pm$\,0.31	& 01:50	\\
20021109  & 13:13\,$\pm$\,14 &  2.99\,$\pm$\,0.32   &    III/II    &  29W  & 13:25\,$\pm$\,01	 & 2.16\,$\pm$\,0.07	& 13:08	\\
20021219  & 21:40\,$\pm$\,04 &  1.74\,$\pm$\,0.10   &    III/II    &  09W  & 21:51\,$\pm$\,01	 & 1.02\,$\pm$\,0.06	& 21:41	\\
20030531  & 02:17\,$\pm$\,05 &  1.85\,$\pm$\,0.13   &    III       &  65W  & 02:20\,$\pm$\,03	 & 1.94\,$\pm$\,0.15	& 02:21	\\
\hline
\end{tabular}
\end{table*}

\begin{table*}
 \renewcommand\thetable{B.1}
\centering \caption{\textbf{Continued.}}
\begin{tabular}{c c c c c c c c c|}
\hline
\hline  \\[-0.5em]
 Event Day& Proton		& Proton Path  	& Proton Rel.	   & Flare	& Electron	& Electron Path		&  Type-III   \\
 	  & Rel. (UT)		& Length (AU)	& Radio Assoc.	   & Long.	& Rel. (UT)	& Length (AU)	&  Start (UT)   \\
\hline   \\[-0.5em]             
20031028  & 10:21\,$\pm$\,07  &  2.21\,$\pm$\,0.16  &    II/[IV-c]      &  08E  & --			& --			& --	\\
20031102  & 09:27\,$\pm$\,05  &  1.91\,$\pm$\,0.16  &    III/II    &  --   & --			& --			& 08:56	\\
20040411  & 04:14\,$\pm$\,06  &  2.14\,$\pm$\,0.11  &    III       &  46W  & 04:15\,$\pm$\,03	& 1.31\,$\pm$\,0.14	& 04:08	\\
20040713  & 00:12\,$\pm$\,09  &  2.77\,$\pm$\,0.21  &    III/II    &  59W  & 00:28\,$\pm$\,03	& 1.28\,$\pm$\,0.12	& 00:14	\\
20040725  & 14:55\,$\pm$\,09  &  2.60\,$\pm$\,0.20  &    III/[IV-c]    &  33W  & --			& --			& 14:45	\\
20041101  & 05:34\,$\pm$\,04  &  2.00\,$\pm$\,0.10  &    III       &  --   & 05:50\,$\pm$\,03	& 1.45\,$\pm$\,0.15	& 05:42	\\
20050115  & 06:23\,$\pm$\,03  &  1.67\,$\pm$\,0.08  &    III/II/[IV-c] &  06E  & 06:23\,$\pm$\,01	& 1.38\,$\pm$\,0.09	& 06:07	\\
20050713  & 14:15\,$\pm$\,08  &  2.77\,$\pm$\,0.18  &    III/[IV-c]    &  80W  & 14:16\,$\pm$\,02	 & 1.27\,$\pm$\,0.12			& 14:02	\\
20050822  & 01:19\,$\pm$\,05  &  1.89\,$\pm$\,0.10  &    III/[IV-c]    &  48W  & -- 		& --			& 01:20	\\
20060706  & 08:27\,$\pm$\,05  &  2.71\,$\pm$\,0.13  &    III/II    &  32W  & 08:37\,$\pm$\,04	& 1.55\,$\pm$\,0.21	& 08:21	\\
20061213  & 02:21\,$\pm$\,11  &  2.13\,$\pm$\,0.41  &    III/II/[IV-c]  &  23W  & --			& --			& 02:24	\\
20100814  & 10:05\,$\pm$\,04  &  1.50\,$\pm$\,0.08  &    III/II/[IV-c] &  54W  & 10:06\,$\pm$\,03	& 1.02\,$\pm$\,0.15	& 10:03	\\
\hline
\end{tabular}
\end{table*}

\end{document}